\documentclass[aps,groupedaddress,nofootinbib,11pt]{revtex4}

\usepackage[dvips]{epsfig}
\usepackage{amsfonts}
\usepackage{amsmath}
\usepackage{}
\usepackage{color}

\definecolor{Blue}{rgb}{0.3,0.3,0.9}
\definecolor{Red}{rgb}{1.0,0.0,0.0}
\definecolor{Green}{rgb}{0,0.4,0}
\definecolor{Violet}{rgb}{0.4,0.0,0.6}
\definecolor{Cyan}{rgb}{0.0,0.4,0.6}
\definecolor{Orange}{rgb}{1.0,0.4,0.0}

\newcommand{\Eqref}[1]{(\ref{#1})}
\newcommand{\ket}[1] {\mbox{$ \vert #1 \rangle $}}

\newcommand{\bra}[1] {\mbox{$ \langle #1 \vert $}}
\newcommand{\abs}[1] {\mbox{$ \vert #1 \vert $}}

\newcommand{\av}[1]{\langle\!\langle \,  #1 \, \rangle\!\rangle}

\newcounter{subequation}[equation] \makeatletter
\expandafter\let\expandafter\reset@font\csname
reset@font\endcsname
\newenvironment{subeqnarray}
  {\arraycolsep1pt
    \def\@eqnnum\stepcounter##1{\stepcounter{subequation}{\reset@font\rm
      (\theequation\alph{subequation})}}\eqnarray}
  {\endeqnarray\stepcounter{equation}}
\makeatother

\newcommand{\ba}{\begin{eqnarray}}
\newcommand{\ea}{\end{eqnarray}}
\newcommand{\sba}{\begin{subeqnarray}}
\newcommand{\sea}{\end{subeqnarray}}

\begin{document}

\vskip 1truecm

 \title{Decoherence and entropy of primordial fluctuations \\
  II. The entropy budget} 
 \vskip 1truecm
 \author{David Campo}
\affiliation{Lehrstuhl f\"{u}r Astronomie,
Universit\"{a}t W\"{u}rzburg, Am Hubland,
D-97074 W\"{u}rzburg, Germany}
 \author{Renaud Parentani}
 \affiliation{Laboratoire
de Physique Th\'{e}orique, CNRS UMR 8627, 
Universit\'{e} Paris-Sud 11, 
91405 Orsay Cedex, France}
 \begin{abstract}
We calculate the entropy of adiabatic perturbations 
associated with a truncation of the hierarchy of Green functions
at the first non trivial level, i.e. in a self-consistent 
Gaussian approximation. 
We give the equation governing the entropy growth
and discuss its phenomenology.
It is parameterized by two model-dependent kernels.
We then examine two particular inflationary models, 
one with isocurvature perturbations, the other with 
corrections due to loops of matter fields.
In the first model
the entropy grows rapidely,
 while in the second  
the state remains pure (at one loop).
\end{abstract}
\maketitle

\section{Introduction}

This is the second part of a series of notes on the decoherence and entropy
of primordial fluctuations predicted by inflation. 
The first part, subsequently called I \cite{paperI}, was 
devoted to the operational formulation of the 
notion of decoherence of metric 
perturbations. Here we turn to the calculation of the entropy
when truncating the hierarchy of Green functions
at the first non-trivial level, first in general
and then in two particular models.
As an introduction, 
we briefly review the present state-of-the-art 
concerning the entropy growth during inflation.

Up to now, the time dependence of the entropy 
as well as the main source(s) of entropy remain undetermined. 
Part of the difficulties arises from the definition of the
reduced density matrix.
Concerning the methodology, previous studies
either work with a master equation for the reduced density matrix 
\cite{Burgess,Martineau}, or directly calculate the 
reduced density matrix 
\cite{MukhaEntropy,Prokopec}, 
or use analogies with other quantum mechanical situations \cite{SPKnew}.

The main problem 
with the first two approaches comes from the  
actual calculation of the trace.
While this poses no difficulty for Gaussian models as in 
\cite{MukhaEntropy,Prokopec},
it is hamperred by the infinities ubiquitous in interacting
Quantum Field Theories. In previous attempts where this calculation was
undertaken \cite{Burgess,Martineau}, 
unjustified assumptions were 
made and lead to contradictory results. 
We argued in paper I that 
the only way to deal properly with these infinities 
is to work with Green functions.
The approach of \cite{SPKnew} is different as it assumes 
a general Gaussian and Markovian {\it Ansatz} 
for the master equation. The limitation of this 
approach is that it does not make 
predictions. Indeed,  
first the parameters of the master equation are undetermined
unless they are calculated in field theoretical settings.
Second, the Markovian hypothesis 
is unjustified for superhorizon 
perturbations at any epoch,  
as well as for perturbations of any scale
during inflation.

In this paper, we present an approach which combines 
the advantages of the previous ones, but without their shortcommings. 
We work with Green functions from the onset.
Infinities are handled 
by the standard renormalization techniques, and the 
general evolution equation 
for the entropy follows straightforwardly
from the knowledge of these 
Green functions.
This emphasis on Green functions is not new \cite{HuBBGKY}, 
but to our knowledge it  
has not been applied to the calculation of the entropy 
of primordial fluctuations in realistic models of inflation.

After presenting the settings in Sec. \ref{sec:Gapprox}, 
the evolution equation of the entropy of the reduced density matrix 
is derived in Sec. \ref{sec:entropygrowth}, followed by a  
general discussion of this equation.
We then examine two specific models.  
A model of two field inflation is presented in Sec. \ref{sec:iso}.
It confirms and generalizes the analysis of \cite{Prokopec}.
In particular we identify the conditions under which 
the entropy grows with the number of efolds.
In Sec. \ref{matterloops} we compute the entropy associated 
to one-loop corrections from matter fields in a class of 
semi-realistic theories analyzed by Weinberg in \cite{Weinberg}.
In these models we show that no significant 
decoherence occurs during inflation.

\section{The Gaussian approximation}
\label{sec:Gapprox}

We work in the coordinate system in which the inflaton field is homogeneous
on each space-like hypersurface, 
\ba
  \varphi(t,{\bf x}) = \varphi_0(t) \, .
\ea
We focus on the curvature perturbations $\zeta$.
In the linear approximation, 
their evolution is described by the action
\ba \label{Squadzeta}
  S = \frac{1}{8\pi G} \int\!\!dt \, d^3x \, a^3 \epsilon 
  \left( \dot \zeta^2 -  
  \frac{1}{a^2} \left( \nabla \zeta \right)^2 
 \right)\, ,
\ea
where $a$ is the scale factor and $\epsilon = - \dot H/H^2$ 
is the first slow roll 
parameter, see \cite{Malda} for a comprehensive derivation.
A dot stands for the derivation w.r.t. the cosmological time $t$.
The Fourier mode labeled with the 
conserved comoving wave vector $\bf q$ obeys
\ba \label{EOMzetafree}
  &&\ddot \zeta_{\bf q} + \frac{d\ln(a^3\epsilon)}{dt}  \dot \zeta_{\bf q} + 
    \frac{q^2}{a^2} \zeta_{\bf q} = 0 \, ,
\ea
where $q $ is the norm of the vector $\bf q$.

The free vacuum is taken to be the Bunch-Davis vacuum,
defined from the positive frequency solutions of \Eqref{EOMzetafree} 
for infinite physical momentum, 
\ba \label{BDmode}
  \left( i\partial_{\tau} - q \right) 
  \left( a\sqrt{\epsilon} \,\zeta_{q}^{\rm in} \right)
  \to 0  \, , \qquad 
  \frac{q}{aH} \to \infty \, .
\ea
$\tau$ is the conformal coordinate time $dt = a d\tau$. 
Using this solution, 
the mode operator can be decomposed as 
\ba
  \zeta_{\bf q}(t) = a_{\bf q}^{\rm BD} \, \zeta_q^{\rm in}(t) 
  + a_{-\bf q}^{\rm BD \, \dagger} \, \zeta_q^{{\rm in}\, *}(t) \, ,
\ea
where the destruction operator $a_{\bf q}^{\rm BD}$ annihilates the Bunch-Davis vacuum.

We shall work in a quasi-de Sitter approximation. 
It can be shown that the results 
are still valid in slow roll inflation. 
In this approximation, the $in$ mode is
\ba \label{solzeta}
  \zeta^{\rm in}_{ q}(t) &=& 
  \zeta_{q}^0 \left( 1+ i q\tau \right) e^{-iq\tau} 
\,  .
\ea
The constant $\zeta_q^0$ is fixed 
by the equal time commutator $[\zeta_{\bf q}, \pi_{\bf q'}^\dagger] = i
\delta^3({\bf q}-{\bf q'})$,
where 
\ba \label{pi_zeta}
   \pi_{\bf q}
   = \frac{a^3 \epsilon}{4\pi G} \, \dot \zeta_{\bf q} 
    \, ,
\ea
is the Fourier component of the momentum conjugate to $\zeta$. 
One finds
\ba \label{norms}
  \abs{\zeta_q^0}^2 = \frac{4\pi G }{\epsilon} \, \frac{H^2}{2 q^3} \, .
\ea

In the linearized description based on Eq. (\ref{Squadzeta}), 
the state of $\zeta$ stays Gaussian and its 
properties are characterized by a single function, 
the power spectrum $\propto \vert \zeta_q \vert^2$ \cite{paperI}.
In interacting field theories, the state is characterized by
the full hierarchy of connected Green functions.
Such knowledge is out-of-reach, so that in practice one resorts 
to a (self-consistent) truncation of this hierarchy. 
This coarse graining  
defines a reduced density matrix with a non vanishing
entropy, see Sec. III in paper I.
The first non trivial level of truncation is defined by 
setting to zero all 
connected Green functions of order $\geq 3$.
The reduced density matrice $\rho_{\rm red}$ is therefore Gaussian.
Since the state is still homogeneous, each
two mode sector $(\zeta_{\bf q},\zeta_{\bf -q})$ can be analyzed separately.
Therefore $\rho^{\rm red}_{{\bf q}, -\bf q}$
is fully characterized~\cite{paperI} by the c-number function
 $G(\tau, \tau'; q)$ given by
\ba
  G(\tau, \tau'; q) \, \delta^3({\bf q}-{\bf q'}) \equiv
   \frac{1}{2} {\rm Tr} \left( \rho^{\rm red}_{{\bf q}, -\bf q}
   \left\{\zeta_{\bf q}(\tau),\, \zeta_{-\bf q'}(\tau') \right\} \right) \, .
\ea 
Since the different sectors do not mix, we shall work at fixed 
$\bf q$ and no longer write the trivial $\delta^3({\bf q}-{\bf q'}) $
coming from the plane wave normalization.

To calculate the entropy carried by $\rho^{\rm red}_{{\bf q}, -\bf q}$ 
it is convenient to recast the information contained 
in $G(\tau, \tau'; q)$ into the covariance matrix $C$ defined by
\ba \label{covmatrix}
    C &\equiv& \frac{1}{2}
  {\rm Tr} \left(  \rho \, \left\{ V ,\, V^{\dagger} 
   \right\} \right)
  =    
\left(
   \begin{array}{cccc}
          {\cal P}_{\zeta} &  {\cal P}_{\zeta \pi} \\
            {\cal P}_{\zeta \pi} &  {\cal P}_{\pi}   
   \end{array} 
   \right) \,, 
 \qquad 
  V =    
\left(
   \begin{array}{c}
          \zeta_{\bf q}\\
          \pi_{-{\bf q}}      
   \end{array} 
   \right) \, .
\ea
As shown in Appendix \ref{app:CTP}, in the Gaussian approximation it is
always possible to make a canonical transformation 
$(\zeta,\pi') \mapsto (\zeta, \pi)$ where the new momentum is 
related to $\dot \zeta$ as in \Eqref{pi_zeta}. 
This canonical transformation leaves invariant the entropy \cite{paperI}.
Using Eq. (\ref{pi_zeta}), 
the three moments ${\cal P}$ are then related to $G$ by 
\sba \label{defrhored}
  {\cal P}_{\zeta}(q,t)  &=& G(t,t;q) 
  \, , 
\\
  {\cal P}_{\zeta \pi}(q,t) 
  &=& \frac{a^3 \epsilon}{4\pi G} \, \partial_{t'}
  G(t,t';q)\vert_{t=t'} 
  \, , 
\\ 
  {\cal P}_{\pi}(q,t) 
  &=& \left( \frac{a^3 \epsilon}{4\pi G} \right)^2 \, 
  \partial_{t}\partial_{t'}
  G(t,t';q) \vert_{t=t'}    
  \, . 
\sea

For the linear perturbations in the Bunch-Davies vacuum, since 
$G(\tau,\tau';q)= \Re \big(\zeta_q^{\rm in}(\tau)\, 
  \zeta_q^{\rm in\, *}(\tau')\big)$
where $\zeta_q^{\rm in}$
is given in Eq. \Eqref{solzeta}, 
the (unperturbed) covariances are 
\sba \label{freecov}
  {\cal P}^0_{\zeta}(q,\tau) &=&  
  \abs{\zeta_q^{\rm in}(\tau)}^2
  = \abs{\zeta_q^0}^2 \left( 1 + x^2 \right)\, ,
\\
  {\cal P}^0_{\zeta \pi}(q,\tau)  &=& \frac{a^3 \epsilon}{4\pi G}
   \, \Re \big(\zeta_q^{\rm in}(\tau)\, 
    \partial_t \zeta_q^{\rm in\, *} \big)
   = -Hx^2 \abs{\zeta_q^0}^2 \left( 
 \frac{a^3 \epsilon}{4\pi G}   \right) \, ,
\\
  {\cal P}^0_{\pi}(q,\tau) &=&
\left(  \frac{a^3 \epsilon}{4\pi G}
 \right)^2 \abs{\partial_t \zeta_q^{\rm in}}^2
  =   H^2 x^4 \abs{\zeta_q^0}^2 \left( 
 \frac{a^3 \epsilon}{4\pi G}  \right)^2
   \, ,
\sea
where we used the relation $\tau \simeq -1/aH$ and introduced 
\ba
   x = \frac{q}{aH} = e^{-N}
   \, .
\ea
$N = - \ln x$ is the number of e-folds with respect to 
horizon exit.

The entropy of $\rho_{{\bf q},\, -{\bf q}}^{\rm red}$ is related to 
the determinant of the covariance matrix by
\ba \label{SvN}
  S &=& 
  2\left[(\bar n + 1 ) \ln(\bar n + 1) - \bar n \ln(\bar n)\right] \, ,
\ea
where the parameter $\bar n$ is defined by
\ba \label{SvN2}
   \left( \bar n + \frac{1}{2} \right)^2 
   \equiv  {\rm det}(C)  = 
   {\cal P}_{\zeta}{\cal P}_{\pi} - {\cal P}_{\zeta \pi}^2 
 \, . 
\ea 
The prefactor $2$ in \Eqref{SvN}
accounts for the fact that $\rho_{({\bf q},\, -{\bf q})}^{\rm red}$ 
is the state of two modes. 
The pure states, $S=\bar n = 0$,  
correspond to $\det(C) = 1/4$. 
At any time, one verifies that 
Eqs. \Eqref{freecov} exactly  
give this constant value of the determinant. Notice also that, 
had we neglected the terms $O(q^2/a^2H^2)$ in (\ref{freecov}a),
we would have instead  
obtained $\det(C) = 0$ which makes no sense in quantum mechanical
settings, since the inequality $\det(C)\geq 1/4$ is nothing but the 
Heisenberg uncertainty relations.

We will also employ another useful parameterization of the 
covariance matrix
\ba
\label{defn}
  n_\zeta(t) &\equiv& {\rm Tr}\left( \rho^{\rm red}_{\bf q,-\bf q}(t) \, 
  a_{\bf q}^{{\rm BD} \, \dagger} a_{\bf q}^{\rm BD}   \right) \, , 
\\
\label{defc}
  c_\zeta(t) &\equiv& {\rm Tr}\left( \rho^{\rm red}_{\bf q,-\bf q}(t)  \, 
  a_{\bf q}^{\rm BD} a_{-\bf q}^{\rm BD}   \right) \, ,
\ea
$n_\zeta$ is real while $c_\zeta$ is complex. 
We recall that these parameters depend on the choice of canonical 
variables, that is on the choice of $(a_{\bf q},a_{\bf q}^\dagger)$,
see \cite{paperI} for more details. So does the parameter 
$\delta_\zeta$ defined in the next equation. 
This parameter  
characterizes the correlations between the modes 
$\bf q$ and $-\bf q$ in the state $\rho_{\rm red}$: 
\ba \label{defdelta}
  \vert c_\zeta \vert^2 \equiv n_\zeta(n_\zeta+1 - \delta_\zeta) \, , \qquad 
  0 \leq \delta_\zeta \leq n_\zeta+1 \, .
\ea 
In this parameterization, the determinant of $C$ is 
\ba \label{detCdelta}
  \det(C) = \left( n_\zeta+\frac{1}{2}   \right)^2 - \vert c_\zeta \vert^2 = 
  \frac{1}{4} + n_\zeta \delta_\zeta \, .
\ea 
The entropy is a monotonously growing function of $\delta_\zeta$. 
This parameter quantifies the residual coherence in the 
state: the state is pure when $\delta_\zeta=0$
and thermal when $\delta_\zeta = n_\zeta+1$. 
The threshold value $\delta_\zeta = 1$ separates entangled states from 
decohered states which cannot be distinguished from statistical 
ensembles \cite{paperI}.
At the threshold value, in the radiation dominated era, the entropy is
\ba \label{Ssep}
  S_{\rm sep} = \ln n_\zeta^{\rm end}
  \simeq - \ln\left(  x_{\rm end}^4\right) 
  = 4 \,  N_{\rm end}
  \, ,
\ea
where 
$N_{\rm end}= -\ln x_{\rm end}$ 
is the number of efolds from horizon 
exit to the end of inflation. 
We have used the linearized treatment to estimate 
$n_\zeta^{\rm end}$ (i.e. $n_\zeta = 1/4x_{\rm end}^4$) \cite{CPfat}.
The maximal (thermal) entropy (per two-modes) is 
$S_{\rm max} = 2 S_{\rm sep}  \simeq 2 \ln n_{\rm end} \simeq
8 N_{\rm end}$.

\section{Evolution of the entropy}
\label{sec:entropygrowth}

\subsection{Equation of evolution}
\label{fundaeqn}
 
As seen from Eqs. \Eqref{SvN} and \Eqref{SvN2}, $S$ is a monotonically
growing function of $\det(C)$. In particular, for $n\delta \gg 1$, 
the entropy is simply given by $S = \ln(n\delta)= \ln (\det C)$. 
Hence, in this regime one has
\ba \label{EOMS}
  n\delta \gg 1  \, , \qquad 
  \dot S =  \frac{d}{dt} \ln(\det C)\, .
\ea
The evolution of $\det(C)$ follows from that 
of the three expectation values of Eqs. \Eqref{defrhored}
which are all determined by the anticommutator $G$. 
Therefore the 
time dependence of $\det(C)$ is governed by the 
equation obeyed by $G$.
The latter is a linear second order 
integro-differential equation. It 
is derived in Appendix \ref{app:CTP}. But, 
as explained in \ref{app:trick}, 
it is simpler to exploit the Gaussianity and 
derive $G$ from an equivalent quantum Langevin equation. 
That is, $G$ is the anticommutator of the operator
$\zeta_{\bf q}(t)$ which verifies the 
effective equation
\ba \label{backreactEOM}  
  \ddot \zeta_{\bf q} + \frac{d\left( \ln(a^3\epsilon)\right)}{dt}  
  \dot \zeta_{\bf q} + \frac{q^2}{a^2} \zeta_{\bf q}(t)  + 
  \int_{-\infty}^{t}\!\!dt' \, D_q(t,t')\zeta_{\bf q}(t')  
  = \xi_{\bf q}(t)\, .
\ea
In the Gaussian approximation, the effects of the interactions are 
summarized in two kernels, namely the ''dissipation'' kernel $D_q$ 
and the so-called noise kernel 
\ba \label{defN}
 N_q(t,t') = \frac{1}{2}\langle \left\{
 \xi_{\bf q}(t), \,  \xi_{-\bf q}(t') \right\}\rangle
 \, .
\ea
$N$ and $D$ appear in the effective action of the curvature perturbation as 
the real and imaginary parts of the renormalized self-energy \Eqref{NetD}.
For the differential part of \Eqref{backreactEOM} we assumed 
the same structure as in the free mode equation \Eqref{EOMzetafree}.
Equation \Eqref{backreactEOM} could be generalized to the case
where the renormalized frequency differs from $q^2/a^2$.  
This would only affect the retarded Green function and $G$
but not the form of the equation \Eqref{EvoldetC}  
governing $\det(C)$.
However our result does not generalize to time dependent 
''wave function renormalization'' since, should they occur, 
they would invalidate the use of Eqs. \Eqref{identity1} and \Eqref{identity2}.

We now proceed assuming that the kernels $D$ and $N$ are known,
and that $G$ and its time derivatives
possess well-defined coincidence point limits.
We derive an equation for the covariances of $\zeta$.
For this we need the identities
\ba 
 \label{identity1}
   \frac{d {\cal P}_{\zeta}}{dt} &=& 2 {\cal P}_{\zeta \dot \zeta} \, ,\\
  \label{identity2}
   \frac{d {\cal P}_{\zeta \dot \zeta}}{dt} &=&  {\cal P}_{\dot \zeta \dot \zeta} + 
   \frac{1}{2} \, 
   \langle \{ \ddot \zeta_{\bf q}(t) ,\, \zeta_{-\bf q}(t) \}  \rangle
   \, .
\ea
The equations can be easily derived in the Heisenberg picture,
since time derivation and taking the expectation value, 
being both linear operations, commute.
These equations are easily verified for \Eqref{freecov} and \Eqref{EOMzetafree}.
Inserting \Eqref{backreactEOM} into \Eqref{identity2} we get
\sba \label{identityEOM}
  {\cal P}_{\dot \zeta \dot \zeta} &=& \dot {\cal P}_{\zeta \dot\zeta} - 
  \frac{1}{2} \langle \{ \ddot \zeta_{\bf q}(t) ,\, \zeta_{-\bf q}(t) \}  \rangle
 \nonumber \\
 &=& \left( \frac{d}{dt} + \frac{d\ln(a^3 \epsilon)}{dt} \right) 
 {\cal P}_{\zeta \dot \zeta}
  + \frac{q^2}{a^2} \, {\cal P}_{\zeta} \,
 \\
 && \, + \, \int_{-\infty}^{t}\!\!dt' \, 
 D_q(t,t')\,  \frac{1}{2} \, 
 \langle  \left\{
 \zeta_{{\bf q}}(t), \,  \zeta_{-\bf q}(t')  \right\}
  \rangle  \, 
 -   \frac{1}{2} 
 \langle  \left\{
  \xi_{\bf q}(t), \,  \zeta_{-{\bf q}}(t)  
  \right\}
 \rangle \, .
\sea
The three terms in (\ref{identityEOM}a) 
are local in time and do not contribute to  
the time evolution of $\det(C)$, as will be seen in Eq. \Eqref{EvoldetC}. 
Instead, the terms in (\ref{identityEOM}b) 
are non local and absent in the free evolution. 
Without them, the time evolution of ${\cal P}_{\dot \zeta \dot \zeta}$
is entirely determined by that of
 ${\cal P}_{\zeta \dot \zeta}$ and ${\cal P}_{\zeta \zeta}$,
 as shown in Eq. (\ref{identity2}).
 
To compute the growth in entropy, we also need 
the time derivative of ${\cal P}_{\dot \zeta \dot \zeta}$, 
obtained from using again \Eqref{backreactEOM},
\ba \label{EOMPdotdot}
 \frac{d{\cal P}_{\dot \zeta \dot \zeta}}{dt} &=&  
  \langle  \{
   \ddot\zeta_{\bf q}, \,  \dot \zeta_{-{\bf q}} \}
   \rangle 
   = - 2 \frac{d \ln(a^3 \epsilon)}{dt} {\cal P}_{\dot \zeta \dot \zeta}
  - 2 \frac{q^2}{a^2} \, {\cal P}_{\zeta \dot \zeta}  
\nonumber \\
  &&\quad \quad  \, - \, \int_{-\infty}^{t}\!\!dt' \, 
  D_q(t,t') \, 
 \langle 
  \left\{
   \dot \zeta_{{\bf q}}(t), \,  \zeta_{-\bf q}(t') 
     \right\}
    \rangle 
+ \,  \langle  \left\{
 \xi_{\bf q}(t), \,  \dot \zeta_{-{\bf q}}(t) 
  \right\}
 \rangle \, .
\ea  
Using the equations \Eqref{pi_zeta}, \Eqref{identityEOM} and \Eqref{EOMPdotdot}, 
we find 
\ba \label{EvoldetC} 
  \frac{1}{2}
  \frac{d\det(C)}{dt} =
  \left( \frac{a^3 \epsilon}{4\pi G} \right)^2
   &&\Big\{
   {\cal P}_{\zeta \dot \zeta}(t)
  \, \int_{-\infty}^{t}\!\!dt' \,   
  \left[ 
 D_q(t,t')\, G(t,t') -   N_q(t,t')\, G_{\rm ret}(t,t') \right] \Big.
\nonumber \\
 && + \, {\cal P}_{\zeta}(t)  
  \int_{-\infty}^{t}\!\!dt' \,
 \left[N_q(t,t') \, \partial_t G_{\rm ret}(t,t') - 
 D_q(t,t')\, \partial_t G(t,t')
 \right] \Big\} \, ,
\qquad
\ea
where $G_{\rm ret}$ is the retarded propagator associated with \Eqref{backreactEOM}.
The prefactor
$(a^3 \epsilon/4\pi G)^2$ comes from the relation \Eqref{pi_zeta}.
We will verify on several examples in Sec. \ref{sec:iso} 
that this prefactor 
yields a rapid growth in $(aH/q)^6$.

Equation \Eqref{EvoldetC} governs the rate of change of the entropy
defined by reducing the state as explained in Section \ref{sec:Gapprox}.
It is {\it generic} and {\it exact}. 
The explicit expressions of the noise and dissipation kernels
are of course model dependent. The important point
is that they can be calculated using standard 
QFT techniques. 

This equation is equivalent to a master equation, but 
has several advantages over it.
The master equation is the evolution equation for the reduced density matrix.
It is obtained by calculating  
the partial trace on the r.h.s. of the Heisenberg equation
\ba \label{master}
  i \frac{d}{dt} \rho_{\bf q,\, -\bf q}^{\rm red} = 
  {\rm Tr}_{{\bf q'}\neq {\bf q}} 
  \, 
  \left( \left[ H ,\, \rho \,  \right]   \right)\, ,
\ea
where $H$ are $\rho$ are the Hamiltonian and density matrix 
of the entire system. The trace is performed on all field
configurations except $\zeta_{\bf q}$.
The difficulty with \Eqref{master} (or the equivalent equation for its
Wigner representation) is to actually calculate the trace.
This is relatively straightforward in non relativistic 
quantum mechanics where the master equation has indeed
proven to be a fruitfull approach, see \cite{decohQM} and 
\cite{Kieferbook,Zurekreview} for recent reviews.
For quantum fields however, 
the first nontrivial contributions to the r.h.s. 
of \Eqref{master} arise from loop corrections. Therefore, the  
calculation 
can only be reliably done at the level of Green functions.
Previous attempts to calculate directly \Eqref{master} 
involved dubious assumptions and approximations 
which have lead to contradictory results, compare e.g. 
\cite{SPKnew,Martineau,Burgess}.
Moreover, Eq. \Eqref{master} does not bring us 
anywhere nearer to the growth of 
entropy since one still has to calculate the covariances. 
In contrast, Eq. \Eqref{EvoldetC} 
involves the renormalized expectations values of the relevant observables.

In the remainder of this section we adopt a
phenomenological approach and discuss how $\dot S$ depends
on the properties of $N$ and $D$. We start with two general remarks. 

As a consistency check we verify that $\det(C)$ is constant 
in an equilibrium state and in a static universe. 
In that case, $N$ and $D$ and $G$ and 
the commutator $G_{[\, \, ]}$
are related by the fluctuation-dissipation (or KMS) relations 
(for each $\bf q$),
\sba \label{FDenv}
  N(\omega) &=& 
  \coth\left( \frac{\omega}{2 k_B T} \right) D(\omega) \, ,
  \\ 
  \label{FDsyst}
  G(\omega) &=& 
  \coth\left( \frac{\omega}{2 k_B T} \right) G_{\rm ret}(\omega) \, ,
\sea 
where $T$ is the temperature.
Inserting these identities into the r.h.s. of \Eqref{EvoldetC} 
yields, as expected, $\det(C) = {\rm cte}$ through a detailed balance.

We also note that out of equilibrium, the sign on the r.h.s. of \Eqref{EvoldetC} 
is a priori undetermined. 
However, one expects on physical grounds that the entropy averaged over 
sufficient time scales does not decrease.\footnote{This is indeed 
what happens, 
for instance, during the early stages of 
the far-from equilibrium evolution of a quantum field \cite{Bergesreview}. Then the 
value of $\det(C)$ oscillates around a monotonously growing mean value.}

We now discuss two limiting cases where $N$ and $D$
are approximately local in time or not. 
We then return to the general discussion of \Eqref{EvoldetC} in light of
these results.

\subsection{Markov approximation}
\label{sec:markov}

This is by definition the regime where the
correlation time of $\xi$ is short compared to the 
characteristic time scale(s) of evolution of the cosmological perturbations. 
In this limit, both $N$ and $D$ are local in time.
Let us assume that the environment resembles a thermal bath
(contrary to the discussion at the end of the previous section 
we do not assume here that $\zeta$ is at equilibrium).
We write the {\it Ansatz}
\ba \label{localNetD}
  N(t,t') = {\cal N}(t) \delta(t-t') \, , \qquad 
  D(t,t') = {\cal D}(t) \frac{\partial}{\partial t'} \delta(t- t')\, ,
\ea
where ${\cal N}(t)$ and ${\cal D}(t)$ are slow functions of time compared
to the correlation time scale of the environment, caused for instance
by the adiabatic expansion of the universe.
According to Eq. \Eqref{backreactEOM} dissipation corresponds 
to ${\cal D} > 0$.
Using the identities 
$G_{\rm ret}(t,t)=0$, $\partial_t G_{\rm ret}\vert_{t'=t} = 1$,
and the definitions \Eqref{defrhored},
Eq. \Eqref{EvoldetC} yields
\ba \label{detCthermal}
  \frac{d\det(C)}{dt} &=&   - 2 {\cal D}(t) \det(C)
  + 2 \left( \frac{a^3 \epsilon}{4\pi G} \right)^2 \, {\cal N}(t)\, {\cal P}_{\zeta}\, .
\ea
It is interesting that in this limit, 
the dissipation kernel acts on the determinant of 
$C$, whereas the noise kernel acts only on the power spectrum times the
factor $a^6$.
To continue the discussion, we simplify Eq. \Eqref{detCthermal}
by dropping the term proportional to $\cal D$. 
This is a bad approximation to describe the approach to equilibrium,
 since \Eqref{detCthermal} would not lead to $\det(C) \to {\rm cte}$.
Far from equilibrium, it leads to an upper bound on the entropy
(since $\cal D$ and $\det(C)$ are both positive).
Then, for $n \gg 1$, the rate of growth of the entropy
per efold is
\ba \label{dotSMarkov}
  \frac{dS}{dN} \simeq \frac{a^6(t) \epsilon^2 H^{-1} {\cal N}(t){\cal P}_{\zeta}(t) }{
  \int^{N}\!\!dN'\, 
  a^6 \epsilon^2  H^{-1} {\cal N}{\cal P}_{\zeta}}\, .
\ea

The relevance of the Markov limit for cosmological perturbations is on the 
contrary disputable.
The typical context where the Markovian limit emerges 
is that of an environment
in a thermal state at high temperature \cite{decohQM}. 
This is irrelevant for inflation 
(both single and multi-fields) since the perturbations are in 
a squeezed state.
This is also unlikely to be a 
realistic model after horizon reentry but before decoupling,
because the scales entering the horizon are not in thermal equilibrium
(since we observe acoustic peaks). We will come back to this point in the conclusions.

\subsection{Non Markovian regime}
\label{sec:heuristic}

At low and vanishing temperatures the Markovian approximation 
fails and all the terms in the r.h.s. of 
\Eqref{EvoldetC} are {\it a priori} of equal importance since $D \sim N$.
Most likely this is the case relevant for (single field) inflation since 
all energy densities associated to particle excitations
are redshifted away (with the exception of multifield inflation
treated in Sec. \ref{sec:iso}).

When $N$ and $D$ are of the same order, we see on \Eqref{EvoldetC}
the possibility of a partial cancellation between the terms proportional to 
${\cal P}_{\zeta}$ and ${\cal P}_{\zeta \dot \zeta}$. 
It is perhaps easier to understand this from the decomposition of the
covariance matrix into the sum of the free contributions 
Eqs. \Eqref{freecov} and  
a remainder $M$
\ba \label{C+M}
  C = C_{\rm free} + M  \, .
\ea
This separation follows naturally from the perturbative calculation
of $G$ in \Eqref{defrhored} where $M$ regroups the 
corrections due to interactions, see Appendix \ref{app:CTP}.  
In the representation (\ref{defn}-\ref{defc}), the decomposition \Eqref{C+M} 
corresponds to 
\ba \label{mapNonAdd}
   n = n_{\rm free} +\alpha \, , \qquad  c = c_{\rm free} + \alpha \, \chi\, .
\ea
where
$\vert c_{\rm free} \vert^2 = n_{\rm free} \left( n_{\rm free} +1 \right)$, 
where $\alpha$ is real and $\chi$ is a complex number with norm smaller than $1$.  
The quantity $\alpha$ is essentially governed by the power 
of the fluctuations of the environment, while $\chi$ 
accounts for a possible squeezing of the state of the environment
(which implies in particular $N(t,t') \neq {\cal N}(t) \delta(t-t')$).
The corresponding value of $\delta$ as defined in \Eqref{defdelta} is
\ba \label{deltaNonStationary}
   \delta(\alpha,\chi) = 2\alpha \left(1 + \frac{1-\alpha}{2n_{\rm free}} \right) 
  \, 
  \left[ 1 - Re\left( \chi e^{-i\arg(c_{\rm free})} \right) \, \right] + 
  \frac{\alpha^2}{n_{\rm free}}(1-\abs{\chi}^2)
   + O\left( \frac{1}{n_{\rm free}^2}\right) \, .
\ea
This equation clearly shows the competition between the contribution $\alpha$, 
that tends to increase the entropy, and the
contribution of $\chi$ when the latter has a phase similar to 
$\arg(c_{\rm free})$. 
In particular, when 
$\arg(\chi) = \arg(c_{\rm free})$, the first term in 
\Eqref{deltaNonStationary} vanishes, $\delta$ is quadratic
 in $\alpha$, and will thus still obey $\delta \ll 1$. 
Equation (\ref{deltaNonStationary}) shows 
that due to subtle interference effects, decoherence may be largely suppressed
when the perturbed state is squeezed as the unperturbed one.
We will have an example of this phenomenon in Sec. \ref{matterloops}
where the noise is due to matter fields in the interacting vacuum.
Given the fine tuned nature of the condition 
$\arg(\chi) = \arg(c_{\rm free})$, it is likely that this is the only 
case where it occurs.
We notice finally that a similar mechanism has long been envisaged 
to circumvent the fundamental limit of quantum noise
(i.e. the shot noise and pressure noise of the laser) in interferometric 
detectors of gravitational waves 
\footnote{In the third generation of those detectors, 
the numerous sources of noise could 
possibly be reduced enough so that the dominant one would be the quantum noise 
of the laser monitoring the position of the mirrors, see for instance \cite{Kimble}. 
In conventional interferometers, the laser produces two types of noise, the 
shot noise proportional to the intensity and the radiation 
pressure noise inversely proportional to
the intensity. The total quantum noise can therefore not be arbitrarily small. 
This minimal value is called the Standard Quantum Limit.  
Noticing that light is squeezed during its travel in the arms of the 
interferometer, Unruh \cite{UnruhGW} proposed 
that judiciously squeezing the light send in 
the so-called dark port of the interferometer could beat the 
Standard Quantum Limit.}.

\subsection{Constant rate of entropy growth}
\label{sec:expdecayingcorrections}

%

One can easily show that the entropy rate depends crucially on
how fast the solution of \Eqref{backreactEOM} asymptotes to a constant
(when it does) outside the horizon.
To do this, we use the equivalence between the quantum and stochastic versions of
Eq. \Eqref{backreactEOM} as explained in Appendix \ref{app:trick}, and 
we consider the {\it Ansatz}  
\ba \label{ansatzzeta}
  \zeta_{q}(t) &=& \zeta_{q}^0 \left( 1+ \alpha x + \beta \frac{x^2}{2}+ 
  \gamma \frac{x^3}{3} + O(x^4) \right)\, .
\ea 
written with the variable $x=q/aH$.
The coefficients $\alpha$,... of the expansion are the sum of two terms,
$\alpha = \bar \alpha + \tilde \alpha$: 
a constant term $\bar \alpha = \av{\alpha}$,
corresponding to the positive frequency solution \Eqref{solzeta} of the
free mode equation, given by
 $\bar \alpha = 0$, $\bar \beta = 1/2$ and $\bar \gamma = i/3$;
and a stochastic component $\tilde \alpha,...$ 
which depend linearly on the stochastic 
source $\xi$ and the dissipation kernel $D$, so that $\av{\tilde \alpha} = 0, ...$.
Of course this {\it Ansatz} does not cover all the cases possible.
We will consider another behaviour below.

From \Eqref{ansatzzeta} we deduce
\ba \label{ansatz}
  \det(C) = \frac{1}{4} 
  \left( \frac{aH}{q}  \right)^4 \left[ 
  c_0 + 2c_1 x + c_2 x^2 + c_3x^3 + O(x^4)   \right]
\ea
where the coefficients are 
\sba
  c_0 &=& \av{\vert \alpha \vert^2 } \\
  c_1 &=&  \av{{\rm  Re}(\alpha^* \beta)} \\
  c_2 &=&  \av{\vert \beta \vert^2 - \frac{\vert \alpha \vert^2}{2}
  + 2\left( {\rm  Re}(\alpha^* \gamma)\right) } - \frac{1}{4} \\
  c_3 &=& \av{2{\rm  Re}(\beta^* \gamma) - 
  \frac{1}{2}{\rm  Re}(\alpha^* \beta) }
\sea 
The following cases may then occur:

\noindent
$\bullet$ $\av{\vert \alpha \vert^2 } \neq 0$, then $c_0 \neq 0$ and $dS/dN = 4$.

\noindent
$\bullet$ $\av{\vert \alpha \vert^2 } = 0$, then $c_0 = c_1 = 0$
(since $\av{\alpha} = 0$).
If in addition $\av{\vert \beta \vert^2} \neq 1/4$, 
then $c_2 \neq 0$ and $dS/dN = 2$.

\noindent
$\bullet$ $\av{\vert \alpha \vert^2 } = 0$ and $\av{\vert \beta \vert^2} = 1/4$, 
then $\beta = \bar \beta$ and $c_0 = c_1 = c_2 = c_3 = 0$. 
Hence $dS/dN = 0$.

\noindent
The threshold of separability, Eq. (\ref{Ssep}), can be reached 
by the end of inflation only in the first case.

Let us consider a second {\it Ansatz} where the conservation of $\zeta$ on 
superhorizon scales is violated by a logarithmic term,
\ba
  \zeta_q(t) = \zeta_q^0 \left( 1 + \alpha \ln(x) + 
  \frac{x^2}{2} + i \frac{x^3}{3} + O(x^4)  \right)\,  ,
\ea
where $\av{\alpha} = 0$.
The covariances are given by
\sba
  {\cal P}_{\zeta\zeta} &=& \vert \zeta_q^0 \vert^2 \left\{ 1 + x^2 + 
  \av{\vert \alpha \vert^2} \ln^2(x) \right\} \, ,
\\
  {\cal P}_{\zeta \dot \zeta} &=& -H \vert \zeta_q^0 \vert^2 
  \left\{ x^2 + \av{\vert \alpha \vert^2} \ln(x) \right\} \, ,
\\
  {\cal P}_{\dot \zeta \dot \zeta} &=& H^2 \vert \zeta_q^0 \vert^2 
  \left\{ x^2 + \av{\vert \alpha \vert^2} \right\}     \, ,
\sea
and the covariance matrix has the determinant
\ba
  \det(C) = \frac{1}{4} + \left( \frac{aH}{q} \right)^6 \av{\vert \alpha \vert^2}
 + O\left( \left( \frac{aH}{q} \right)^4 \ln\left( \frac{aH}{q} \right) \right)\, .
\ea
The entropy grows with the rate $dS/dN \simeq 6$.
Higher rates necessitate a stronger violation of the 
constancy of $\zeta$. This behaviour will be observed below
in multifield inflation.

\section{Coupling to isocurvature perturbations during multifield inflation}
\label{sec:iso}

In multifield inflation, 
adiabatic and isocurvature linear perturbations 
can be coupled on scales larger than the Hubble radius.
As a result, even after their decay, 
isocurvature perturbations have affected the primordial 
curvature spectra in an irreversible way.
In this case, tracing over the isocurvature perturbations furnishes 
a non zero entropy at tree level.
This source of entropy was considered in \cite{Prokopec}. 
We generalize the analysis to a wider class of models 
while using the method of Section \ref{sec:entropygrowth}.
We clarify the conditions leading to a linear growth of the 
entropy with the number of efolds. 
In particular we consider
the (non-intuitive) limiting case where 
the curvature of the
background trajectory in field space is very weak.

\subsection{The model}
 
This subsection contains review material and may 
be skipped by the learned reader.
We consider the following class of two-field models \cite{multifield}
\ba \label{actioniso}
  S = \int\!\!d^4x \sqrt{-g} \left[ \frac{R}{16\pi G} - 
  \frac{1}{2} \left( \partial \varphi \right)^2 - 
  \frac{e^{2b(\varphi)}}{2} \left( \partial \chi \right)^2 -V(\varphi, \chi) 
  \right]\, ,
\ea
with a non standard kinetic term for the field $\chi$, 
thereby generalizing the action considered in \cite{Prokopec}.
The Klein-Gordon equations for the homogeneous background fields are
\ba \label{backgroundKG}
  &&\ddot{\varphi} + 3H \dot \varphi + V_{, \varphi} =
   b_{, \varphi} e^{2b} \dot \chi^2
\, , \qquad   
  \ddot{\chi} + (3H + 2b_{,\varphi} \dot \varphi) \dot \chi +
   e^{-2b}V_{, \chi} = 0\, ,
\ea
where the subscripts $_{,i}$ designate a partial
derivative with respect to the field indicated. 
The Einstein equations are
\ba
   H^2 &=& \frac{8\pi G}{3} \left[ \frac{\dot \sigma^2}{2} + V  \right] 
\, , \qquad 
  \dot H = - 4\pi G \dot \sigma^2 \, ,
\ea
where we introduced the field
\ba
  \dot \sigma^2 \equiv \dot \varphi^2 + e^{2b} \dot \chi^2\, .
\ea
The linear cosmological perturbations are perhaps most transparently written in 
the instantaneous basis $\delta \sigma$ and $\delta s$ 
of perturbations respectively tangent and orthogonal to the background trajectory
in the field space $(\varphi,\chi)$ \cite{Wands}
\ba
   \delta \sigma \equiv \cos(\theta) \delta \varphi + 
   \sin(\theta) e^{b} \delta \chi 
\, , \qquad 
   \delta s \equiv -\sin(\theta) \delta \varphi + \cos(\theta) e^{b}
    \delta \chi  \, , 
\ea
where 
\ba \label{theta}
   \cos\theta \equiv \frac{\dot \varphi}{\dot \sigma} \, , \qquad 
   \sin\theta \equiv \frac{e^{b}\dot \chi}{\dot \sigma} \, .
\ea
$\delta \sigma$ is the adiabatic component of the 
vector of linear perturbations, and 
$\delta s$ is called the entropy component. 
In the present class of models, the 
anisotropic stress vanishes and the line element in the longitudinal gauge
simplifies to 
\ba
  ds^2 = - \left(1 + 2\Phi  \right)dt^2 + 
  a^2\left( 1 - 2\Phi  \right) \delta_{ij}dx^i dx^j\, ,
\ea
where $\Phi$ is the gravitational potential.
The gauge invariant curvature perturbation (in the comoving gauge) is
\ba \label{defzeta}
  \zeta &\equiv& \Phi - \frac{H}{\dot H} \left( \dot \Phi + H \Phi \right)
=  \Phi + \frac{H}{\dot \sigma} \delta \sigma\, ,
\ea
where the second expression is obtained using the perturbed 
 momentum constraint equation
$\dot \Phi + H \Phi = 4\pi G \dot \sigma \delta \sigma$. 
One also introduces the dimensionless isocurvature perturbation during inflation
\ba
  {\cal S} \equiv \frac{H}{\dot \sigma} \delta s\, ,
\ea
which is gauge invariant by construction.

The linearized equation for $\zeta$ is remarkably simple,
\ba \label{EOMzeta+iso}
  \ddot \zeta + \frac{d\ln(a^3 \epsilon)}{dt} \dot \zeta + \frac{q^2}{a^2} \zeta = 
  \frac{1}{a^3\epsilon} \frac{d}{dt}\left( -2 a^3 \epsilon \, 
  \frac{V_{,s}}{\dot \sigma} {\cal S}  \right)  \equiv \xi(t)\, ,
\ea
where $\epsilon = - \dot H / H^2$.
The equation for the isocurvature perturbations is a little bit more involved
\ba \label{EOMiso+zeta}
   \ddot {\cal S} + \frac{d\ln(a^3 \epsilon)}{dt} \dot {\cal S} + 
 \left[ \frac{q^2}{a^2} + {\cal C}_{\cal S \cal S}  
  \right]{\cal S} = - 2 V_{,s} \left(\frac{q}{aH} \right)^2 
   \frac{\Phi}{\epsilon}\, .
\ea
The time dependent function ${\cal C}_{\cal S \cal S}$,
\ba \label{Css}
  {\cal C}_{\cal S \cal S} &=& V_{,ss} + 3\dot \theta^2 - 
  b_{, \varphi\varphi} \dot \sigma^2 + b_{,\varphi}^2 g(t) + b_{,\varphi} f(t)
  + \frac{3}{2}H \frac{\dot \epsilon}{\epsilon} + 
 \frac{1}{2}\frac{\ddot \epsilon}{\epsilon} - 
 \frac{1}{4}\left(\frac{\dot \epsilon}{\epsilon} \right)^2
\, , \nonumber \\
  f(t) &=& V_{,\varphi}\left( 1+\sin^2 \theta\right) - 4 V_{,s} \sin\theta 
\, , \nonumber \\
  g(t) &=& - \dot \sigma^2\left( 1+ 3\sin^2 \theta\right)\, , 
\ea
depends only on background quantities.
Finally, $\zeta$ and ${\cal S}$ are coupled via the function
\ba
  g &\equiv& - 2 \frac{V_{,s}}{H\dot \sigma} 
  = \frac{2}{H} 
   \left( \dot \theta + b_{,\varphi} \dot \sigma \sin(\theta)  \right)
\nonumber \\
  &\simeq& 2\eta_{s\sigma} - 2b_{,\varphi} \frac{\dot \sigma}{H} \sin^3(\theta)\, .
\ea
The last expression is valid in the slow-roll approximation, and 
$\eta_{s\sigma}$ is the slope parameter of the potential 
\ba
  \eta_{s\sigma} \equiv \frac{V_{,s\sigma}}{3H^2}\, .
\ea
In brief, 
$\zeta$ and ${\cal S}$ are correlated when
the background trajectory is curved (in field space).

\subsection{Qualitative discussion}
\label{sec:strongcoupling}

Equations \Eqref{EOMzeta+iso} and \Eqref{EOMiso+zeta}
are sufficient 
to understand qualitatively the possible behaviours of the entropy.
These equations differ in two qualitative ways, through their effective mass 
and the source.
Let us first consider the r.h.s. of the equations. 
The source $\xi$ of the adiabatic perturbations is a linear combination of 
${\cal S}$ and its time derivative, but does not
contain spatial derivatives. It cannot in general be neglected
on superhorizon scales.
The source of ${\cal S}$ on the contrary is second order in the gradiant
of $\Phi$ and decays exponentially fast. It can therefore be neglected
in the long-wavelength approximation.

Let us now examine the homogeneous equations.
While $\zeta$ is massless, and therefore reaches a 
constant value outside the horizon,
isocurvature perturbations 
tend to decay because of the effective mass $C_{{\cal S}{\cal S}}$.
More precisely, on superhorizon scales (in the slow roll approximation),
equation \Eqref{EOMiso+zeta} gives
\ba
   {\cal P}_{\cal S} = {\cal P}_{\cal S}^0 
   \left(\frac{H}{H_*}\right)^{-2{\rm Re}(\nu)}
   \left(\frac{q}{aH}\right)^{3-2{\rm Re}(\nu)}\, , 
\ea
where 
\ba 
   \nu^2 = \frac{9}{4} - \frac{{\cal C}_{\cal S \cal S}}{H^2}\, .
\ea
For ${\cal C}_{\cal S \cal S} > 9 H^2/4$, 
${\cal P}_{\cal S}$ decreases like $1/a^3$ 
which is the damping factor of a massive field.

The entropy gained by $\zeta$ after tracing over ${\cal S}$  
follows therefore one of two behaviours.
If $C_{{\cal S}{\cal S}} \geq 9 H^2/4$, the isocurvature component 
remains unexcited (no parametric amplification). 
The entropy gain is therefore negligeable, frozen at its value 
at horizon exit, 
\ba \label{Siso1}
  S\left(\frac{q}{aH} \ll 1 \right) \simeq S\left(\frac{q}{aH} = 1 \right) 
  \equiv S_* \, .
\ea 
We have not attempted to calculate this value. 
In general, is depends on the integrated effect of $\epsilon$, $b_{,\varphi}$, 
and the slope parameters $\eta_{ij}$. 
The analysis of the last reference 
in \cite{multifield} shows that in a wide class of models, 
the slow roll approximation yields good estimates of the power
spectra at horizon exit. In this class, 
they depend only on the values of
the parameters at horizon exit. Unless some fine tuning of these 
parameters, we expect the amplitude of $S_*$ to be 
\ba
  S_* = O(\eta_{s\sigma}^2, \epsilon^2, b_{,\varphi}^2)\, .
\ea
In this case, the reduced state of $\zeta$ remains 
quantum mechanically entangled 
since $S_* \ll S_{\rm sep}$, see Eq. (\ref{Ssep}).

The other case where $C_{{\cal S}{\cal S}} < 9 H^2/4$ is much
more interesting.
Below we now show that the entropy grows linearly with the number of 
efolds as long as the product
$g {\cal S}$ decays slower than $1/a^3$.

\subsection{Entropy growth when isocurvature modes are excited}

The analysis simplifies considerably in the long wavelength limit as 
the damping kernel is negligeable.
Indeed, in order to get the equation \Eqref{backreactEOM} for this model,
we write the solution of \Eqref{EOMiso+zeta} as 
\ba \label{soliso}
  {\cal S}(t) = {\cal S}_h(t)  
  -2 \int_{-\infty}^{t}\!\!dt' \, G_{\rm ret}^{\cal S}(t,t')\, 
  \left(V_{,s} \left(\frac{q}{aH} \right)^2 
   \frac{\Phi}{\epsilon} \right)\, ,
\ea
$G_{\rm ret}^{\cal S}$ is the retarded Green function 
of \Eqref{EOMiso+zeta}.
${\cal S}_h$ is solution of the homogeneous equation, fixed by choosing 
the adiabatic vacuum for $q/aH \gg 1$.
Substituting this solution into \Eqref{EOMzeta+iso}, one has
\ba \label{effEOMzeta}
  \ddot \zeta + \frac{d\ln(a^3 \epsilon)}{dt} \dot \zeta + \frac{q^2}{a^2} \zeta 
  - \frac{4}{a^3\epsilon} \frac{d}{dt}\left(
 \frac{ a^3 \epsilon \, V_{,s}}{\dot \sigma}
  \int_{-\infty}^{t}\!\!dt' \, G_{\rm ret}^{\cal S}(t,t')\, 
  \left(V_{,s} \left(\frac{q}{aH} \right)^2 
   \frac{\Phi}{\epsilon} \right)  \right) 
  = \xi(t) \qquad 
\ea
where $\xi$ is given in \Eqref{EOMzeta+iso} 
with the substitution ${\cal S} \mapsto {\cal S}_h$.
By identification with \Eqref{backreactEOM}, we see that $D$ is
suppressed by a factor $\left(\frac{q}{aH} \right)^2$.
We neglect it in the following with the resulting simplification that
\Eqref{effEOMzeta} reduces to \Eqref{EOMzeta+iso}. Therefore
 all the expressions below are only
valid for $q/aH \ll 1$.

The solution of \Eqref{EOMzeta+iso} has the following form.
Before horizon exit, it is given by Eq. \Eqref{solzeta}.
Then it is given by
\ba  \label{sol2}
  \zeta(t) &=& \zeta^{\rm in}(t) + 
  \int_{t^*}^{t}\!\!dt' \, G_{\rm ret}(t,t') \xi(t') \, ,
\ea
where $\zeta^{\rm in}(t)$ is given by \Eqref{solzeta} and $G_{\rm ret}$ is 
the retarded Green function of \Eqref{EOMzetafree}.
To evaluate the integral of \Eqref{sol2}, it is sufficient 
to use the long wavelength approximation 
\Eqref{Gret} of the retarded Green function. 
More detailed expressions are given in Appendix \ref{app:xi-model}.

To estimate qualitatively the impact of the source $\xi$,
we first consider a simplified model 
where $\xi$ is constant during $N_\xi$ 
e-folds after horizon exit and where it vanishes afterwards. 
This assumption may appear unrealistic at first, but
in the light of the analytical solution of the next section,
we will see that it contains the essential physics
 to understand the evolution of the entropy.
We therefore consider the source term
\ba \label{ximodel}
   \xi(t) = \theta(t_{\xi} - t )\theta(t - t_* )\,  \xi \, ,
\ea
where the constant value $\xi$ is given by
\ba
  \xi \simeq  3H^2 g \, {\cal S}_0 \, .
\ea
Then, for $t_* \leq t \leq t_\xi$, 
$\zeta$ of \Eqref{sol2} tends to 
\ba
  \zeta(t) &=& \zeta^{\rm in}(t) 
  + g {\cal S}_0 \ln\left(\frac{a}{a_*} \right)\, ,
\nonumber\\
  \dot \zeta(t) &=&  \dot \zeta^{\rm in}(t) + g H {\cal S}_0 \, .
\ea 
(Notice that this could have been directly derived from 
the evolution equation 
\ba \label{EOMcompact}
  \frac{d\zeta}{dN} = 
  - \left(\frac{q}{aH} \right)^2 \frac{ \Phi }{\epsilon}+ g{\cal S}
\ea
assuming ${\cal S} = {\rm cte} = {\cal S}_0$ 
and neglecting of the first term.)

Taking into account that at horizon exit, the power spectra are 
\ba \label{powerexit}
  {\cal P}_{\cal S}^0 \simeq {\cal P}_{\zeta}^0 \simeq 
   \frac{4\pi G}{\epsilon} \frac{H_*^2}{2q^3} \, ,
\ea
the covariances for $t_\xi \geq t \gg t_*$ are given by
\sba \label{C11iso}
  {\cal P}_{\zeta} &=& {\cal P}_{\zeta}^0 
  \left[ 1 + g^2 \ln^2\left(x \right) \right]\, ,
\\
\label{C12iso}
  {\cal P}_{\zeta \dot \zeta} &=& 
 {\cal P}_{\zeta \dot \zeta}^0 - 
    g^2 H {\cal P}_{\zeta}^0 \ln\left(x \right) \, ,
\\
\label{C22iso}
  {\cal P}_{\dot \zeta \dot \zeta} &=&
  {\cal P}_{\dot \zeta \dot \zeta}^0 + 
 g^2 H^2 {\cal P}_{\zeta}^0 \, .
\sea 
Recalling the relation \Eqref{pi_zeta} between $\pi$ and $\dot \zeta$, we get
\ba \label{deta^6}
  \det(C) \simeq 
  \frac{1}{4} + g^2 \left( \frac{aH}{q} \right)^6 \,\left[ 
  1 + O\left(g^2 x^2,\,  
  g^2 \ln^2\left(x \right) \right) \right]\, .
\ea
We left the subdominant $1/4$ to remind that $\det(C) \geq 1/4$
by Heisenberg uncertainty relations.
Hence, within the interval
\ba
  H t_* 
   + \vert \ln 4g^2 \vert
   \leq H t \leq H t_\xi  \, ,
 \label{dom}
 \ea
 the entropy grows with a rate
\ba
   \frac{dS}{dN} = 6 \, . \label{highrate}
\ea
This expression is rather remarkable since it is
independent of the parameters of the model and of the initial
conditions of $\varphi$ and $\chi$. These 
appear in the expression of the entropy only as logarithmic 
additive constant, e.g. $2\ln(g)$, and in the boundary of the
domain of (\ref{dom}).
These features seem to be a generic property of  
classically unstable systems.

After the decoupling of curvature and isocurvature perturbations, 
we verify that the entropy is again constant.
The expressions are given in Appendix \ref{app:longlambda}.

 \subsection{Canonical kinetic term}
\label{sec:cankinterm}

To confirm the physical relevance of the above result, we 
study in more details the dependence of the entropy on $g$ and 
on the properties of the background trajectory. 
To this end, we turn to the case 
\ba
  b_{,\,\varphi} = 0 \, , \qquad 
  V = \frac{1}{2}\left( m_{\chi}^2 \chi^2 + m_{\varphi}^2 \varphi^2  \right)\, ,
\ea 
for which analytical solutions exist in the slow roll 
approximation \cite{IsoStaro}.
We shall also explain the counter-intuitive 
result of \cite{Prokopec},
namely that in the limit of a small mass difference, the entropy
grows linearly till the end of inflation 
with the rate \Eqref{highrate}
while the background trajectory is almost straight.

In the slow roll regime, 
the background trajectory can be written in the parametric form
\ba
  \chi = 2M_{\rm Pl} \sqrt{s} \sin \alpha \, , \qquad
  \varphi = 2M_{\rm Pl} \sqrt{s} \cos \alpha \, , \qquad
 0\leq \alpha < \frac{\pi}{2}\, ,
\ea
where 
\ba
   s = \ln\left(\frac{a_{\rm end}}{a} \right) 
   = - \ln\left(\frac{x_{\rm end}}{x} \right)  \, . \qquad 
\ea
Forward time propagation corresponds to decreasing values of $s$.
The slow roll regime corresponds to $s \gg 1$ and ends at $s \simeq 1$.
We call $\chi$ the heavy field and introduce the relative mass
difference 
\ba
   \lambda = \frac{m^2_{\chi} - m^2_{\varphi}}{m^2_{\varphi}} > 0\, .
\ea
The evolution equation
\ba \label{varalpha}
  \frac{d\alpha}{d\ln s} = 
  \frac{\lambda}{4} \frac{\sin 2\alpha}{1 + \lambda \sin^2 \alpha}\, ,
\ea
is obtained from the Klein-Gordon equations \Eqref{backgroundKG}. 
Its solution can be written 
\ba
 s = s_0 \left( \frac{t}{t_0} \right)^{\frac{2}{\lambda}} 
    \frac{1+t^2}{1+t_0^2}  
   = \bar s \,  t^{\frac{2}{\lambda}}\left( 1+t^2 \right) 
 \, , 
\ea
where $t\equiv \tan(\alpha)$.
The index $0$ refers to the initial conditions.
From the solution written in this form we immediately see that
$\alpha$ is a monotonously growing function of $s$. 
Omitting the contribution of $\dot \sigma^2 /2$, 
the Hubble parameter is in turn given by
\ba \label{exactHubble}
  H^2(s) = \frac{2}{3}m^2_{\varphi}s \left( 1 + \lambda \sin^2 \alpha \right)\, .
\ea

The coupling parameter $g$ has the form
\ba \label{exactg}
  g = - 2 \frac{d \theta}{ds} =
  -\frac{\lambda}{3} \frac{m^2_{\varphi}}{H^2} \sin 2\theta\, ,
\ea
where $\theta$ and $\alpha$ are related by
\ba \label{theta-alpha}
  \tan\theta = \tan\alpha 
  \frac{1+ 2\frac{d\alpha}{d\ln s} \frac{1}{\tan \alpha}}{
  1 - 2\frac{d\alpha}{d\ln s} \tan \alpha} \, .
\ea
Substituting \Eqref{varalpha} into \Eqref{theta-alpha} yields
\ba \label{theta-alpha2}
  \tan \theta = (1+\lambda ) \tan\alpha\, ,
\ea
which is exact in the slow-roll approximation.
In the limit of small mass difference $\lambda \ll 1$, 
the velocity vector is almost aligned with the position vector,
hence the background trajectory is almost straight in this case.

After horizon exit, 
the isocurvature perturbations are given by
\ba \label{exactiso}
  {\cal S} &=& {\cal S}_0  \frac{H^2}{H^2_*} \frac{\sin 2\theta}{\sin 2\theta_*}
\equiv {\cal S}_0 {\cal T}_{\cal S}(s,s_*) \, ,
\ea
where ${\cal S}_0$ is as previously the amplitude at horizon crossing
given in the first approximation by \Eqref{powerexit}.
The gravitational potential is determined by the constraint
$\dot \Phi + H \Phi = 4\pi G \dot \sigma \delta \sigma$.
Together with \Eqref{defzeta} it implies that
$\zeta$ still reaches a constant and that
\ba
  \frac{\Phi}{\epsilon} \simeq  \zeta \left( 1+O(\epsilon) 
   \right)  \, .
\ea
Inserting this solution into \Eqref{EOMcompact}, one gets
\ba \label{EOMcompact2}
   \frac{d\zeta}{ds} = 
  \left(\frac{q}{aH} \right)^2 \zeta - g{\cal S} \, .
\ea 
The first term on the r.h.s. can therefore be neglected. 
Since $g {\cal S}$ decays like a 
power of $s$, which is integrable at $s=0$,
the amplitude $\zeta(s)$ indeed asymptotes to a constant.
In the limit $\lambda \ll 1$, integrating equation \Eqref{EOMcompact2} yields
\ba
  \zeta(s) - \zeta_0 &=& {\cal S}_0 {\cal F}(s,s_*) = 
  - 2 {\cal S}_0
  \, \int_{\theta_*}^{\theta(s)}\!\!d\theta \, 
  {\cal T}_{\cal S}(s(\theta),s_*)  \,,
\ea
where $\zeta_0$ is the amplitude at horizon exit and
we used \Eqref{exactg} and \Eqref{exactiso}.
This integral is manifestly convergent but its expression will not be needed here.
Therefore the three covariances are given by
\sba
{\cal P}_{\zeta\zeta} &=& {\cal P}^0 \left( 1 + x^2 +  {\cal F}^2(s)  \right)
 \, ,\\ {\cal P}_{\zeta \dot \zeta} &=& {\cal P}^0_{\zeta \dot \zeta}  + 
  {\cal P}_0 \,  g{\cal T}_{\cal S} {\cal F} 
\, ,\\
  {\cal P}_{\dot \zeta \dot \zeta} &=& {\cal P}^0_{\dot \zeta \dot \zeta}  + 
  {\cal P}_0 \left( g{\cal T}_{\cal S}\right)^2  \,  .
\sea
${\cal P}^0_{\zeta \dot \zeta}$ and ${\cal P}^0_{\dot \zeta \dot \zeta}$ 
are given by Eqs. (\ref{freecov}b) and (c).
The part of $\dot \zeta$ driven by ${\cal S}$ decays polynomially
in $s$ whereas the homogeneous solution decays as $(s/s_*)^2\exp(2(s-s_*))$.
In consequence, in the determinant the term 
${\cal P}_0^2  \left( g{\cal T}_{\cal S}\right)^2 $ 
rapidely overtakes the others, and we have
\ba \label{universal}
  \det(C) \simeq \frac{1}{4}\left[ 1 + 
  \left(\frac{aH}{q}\right)^6 \left( g{\cal T_{\cal S}}\right)^2 \right]\, .
\ea
where $g{\cal T}_{\cal S}$ depends polynomially on $s$.
Hence as in the previous section, we find that the entropy depends 
only logarithmically on 
parameters and initial conditions of the theory via 
$\ln g_*$ and $\ln s_0$.
In other words, $g$ needs to be exponentially small in order to  
suppress the growth of
the factor $\left( aH/q \right)^6$.

More puzzling is the counter-intuitive result~\cite{Prokopec} 
implying that the smaller $\lambda$,
the smaller the coupling $g$ but 
the higher the final value of the entropy because  
 the longer it grows.
We can understand this qualitatively in the following way.
Examination of \Eqref{Css} shows that in the slow-roll approximation,
the leading contribution to the mass of ${\cal S}$ is the curvature of the potential,
\ba
  V_{,ss} = m_{\varphi}^2 \left(1+\lambda \cos^2\theta  \right) \leq 
  m_{\varphi}^2 \left(1+\lambda  \right)\, .
\ea
The Hubble rate is such that $3H^2/2 \geq m_\varphi^2 s$, where the 
slow-roll condition requires $s\gg 1$.
Combining these two inequalities, we get
\ba \label{nicebound}
  \frac{4C_{\cal S \cal S}}{9H^2} \simeq \frac{4V_{,ss}}{9H^2} 
  \leq \frac{2}{3} \, \frac{1+\lambda}{s}\, .
\ea 
Recalling that isocurvature perturbations decay slowlier than $a^{-3}$ when 
$4C_{\cal S \cal S}/9H^2 < 1$, we see that 
the smaller $\lambda$, the easier it is to satisfy this condition.
For $\lambda$ sufficiently small it may be verified until the 
end of the slow-roll regime.
On the contrary, for $\lambda \gg 1$, ${\cal S}$ ceases to be a 
source of entropy $O(\lambda)$ efolds before the end of inflation.
In the opposite regime of extremely weak coupling, $\lambda \ll 1$,
$g \propto \lambda$ only enters in a logarithmic shift of the time
lapse where (\ref{highrate}) is valid, see (\ref{dom}).

\subsection{Discussion} 
\label{sec:discussIso}

A high entropy growth rate
and the logarithmic dependence on the parameters implied by 
$S = \ln \det(C)$ and \Eqref{universal}
seem to be generic for systems whose equations of motion are
characterized by an instability. 
For cosmological perturbations, the 
instability is the mechanism of parametric amplification itself.
Recall indeed that $a^6 = a^3 \times a^3$, where each factor $a^3$ is brought 
by the relation \Eqref{pi_zeta} between $\pi$ and $\dot \zeta$. 
When $\zeta$ is decoupled, two 
factors of $1/a^3$ coming from the Wronskian of $\zeta$ cancel this 
$a^6$ thereby giving $\det(C) = {\rm cte}$. 
Instead, in the presence of a source $\xi$,
the change in $\det(C)$ is given by $a^6$ times 
$({\cal P}_{\zeta}^0 \delta {\cal P}_{\dot \zeta \dot \zeta} 
 + {\cal P}^0_{\dot \zeta \dot \zeta} \delta{\cal P}_{\zeta} - 
 2 {\cal P}^0_{\zeta \dot \zeta}\delta {\cal P}_{\zeta \dot \zeta})$.
This product rapidely asymptotes to
$a^6 {\cal P}_{\zeta}^0 \delta {\cal P}_{\dot \zeta \dot \zeta}$
unless the source $\xi$ decays faster than $1/a^3$.
In other words, the conditions of an efficient decoherence are provided
by the same mechanism producing the highly non-classical state
in the first place.

When the growth of entropy follows Eq. (\ref{highrate})
till the end of inflation, the quantum coherence of the reduced 
state is lost since the resulting value of the entropy is
$S_{\rm end}\sim 6 \ln N_{\rm end}$ which is higher than the 
threshold of separability given in Eq. (\ref{Ssep}). 
We recall that this large entropy is however
compatible with the classical coherence of the
distribution (which implies the presence of the acoustic peaks)
because it is much smaller
than the thermal entropy $= 8 \ln N_{\rm end}$
which characterizes the incoherent distribution (with no peak)
\cite{SPKentropy}.

In the model studied in Sec. \ref{sec:cankinterm},
the law \Eqref{highrate} is 'generic' in that
lower rates $0< dS / dN < 6$ are only found when 
$4C_{\cal S \cal S}/9H^2$ approaches one
before the end of inflation.
Once $4C_{\cal S \cal S}/9H^2 \geq 1$, the entropy saturates at
$S \simeq S_* + 6 N_\xi$,
with $S_* = O(\eta_{s\sigma}^2, \epsilon^2, b_{,\varphi}^2)$
and $N_\xi$ is the number of e-folds from horizon exit till that moment.
Hence the threshold of separability is reached when
$ N_\xi \geq \frac{2}{3}  N_{\rm end} $. 
From the inequality \Eqref{nicebound}, this translates into an 
upper bound on
the relative mass difference $\lambda$. 
To get a rough estimate of this bound, we simplify 
$4C_{\cal S \cal S}/9H^2$ by its upper bound \Eqref{nicebound}.
The condition $2(1+\lambda)/3s_\xi = 1$ with 
$s_\xi = N_{\rm end} - N_\xi$
then gives the upper bound
\ba
   \frac{N_{\rm end}}{2} \,  e^{- 3 N_\xi} \leq \lambda 
   \leq \frac{N_{\rm end}}{2} - 1\, . 
\ea
The lower bound on $\lambda$ comes from (\ref{dom}) and 
$g_* \sim \lambda /s_* = \lambda /N_{\rm end}$. It is easily
satisfied when the number of e-folds $N_\xi $ is not fine tuned 
(e.g. by the choice of initial conditions) to be
 small.
 
 It is simple to extend our conclusion to inflationary models 
 with $\cal N$ fields. Then, $\xi$ is the sum of the 
 contributions of the ${\cal N} - 1$ isocurvature modes \cite{multifieldpertrurb}.
 Hence, the entropy grows steadily 
 as long as at least one isocurvature mode
 is excited.

It is harder to reach conclusions in the case of 
multifield inflation with a non-canonical term. 
In this case, the derivatives of 
$b(\varphi)$ can contribute significantly to $C_{\cal S \cal S}$
so as to increase or lower the effective mass.
Moreover, the coupling can be strong enough to 
lead to a significant change of $\zeta$ on superhorizon scales
and to invalidate a perturbative treatment.
A numerical treatment is probably required to settle the 
question.

\section{Coupling to matter fields} 
\label{matterloops}

In a non stationary background, 
there are {\it a priori} two distinct sources of entropy from loop corrections.
The first is dissipation, as in non vacuum states 
in Minkowski space.
The second is a possible non trivial time dependence of 
radiative corrections
that would frustrate the cancellation between the variances 
which at tree level lead to $\det(C) =1/4$.
Weinberg \cite{Weinberg} identified the conditions such that the corrections 
depend only on the values of background quantities at horizon crossing
(instead of the whole history after horizon crossing).
These conditions are satisfied by a wide class of semi-realistic theories 
\cite{Weinberg2,WeinbergStudent}.
We verify that in these theories, the entropy vanishes at one loop
approximation.

\subsection{The model}

We consider theories of one inflaton field $\varphi(t,{\bf x})$. 
Matter is modeled by ${\cal N}$ copies of free scalar field $\sigma$.
More precisely, we consider the vector 
field $\vec \sigma(t,{\bf x}) = \left( \sigma_1,\, ... , \sigma_{\cal N} \right)$ 
in the fundamental representation of $O({\cal N})$. 
We assume no symmetry breaking otherwise
isocurvature perturbation are excited. 
The gravitational sector was described in Sec. \ref{sec:Gapprox}.
The gravitons have a similar action but as explained below we will not need 
to take them into account.
The quadratic part of the action describing the free evolution of the 
matter fields is
\ba
  S[\sigma_n] = \frac{1}{2}\sum_{n=1}^{{\cal N}} \int\!\!dt d^3x \, 
  a^3 \left[ \dot \sigma_n^2 - \frac{1}{a^2} \left( \nabla \sigma_n \right)^2 
  - 12 \xi H^2 \sigma_n^2   \right]
\ea
Minimal (conformal) coupling to gravity corresponds to $\xi = 0$ 
(resp. $\xi=1/6$).

$\zeta$ and the matter fields are canonically quantized. We work in the 
interacting picture.
The momenta conjugate to $\sigma_n$ are
\ba \label{pi_sigmafree}
   \pi_i(t,{\bf x}) = a^3 \, \dot \sigma_i(t,{\bf x})
\ea
and the mode equations are
\ba  \label{EOMsigma}
  &&\ddot \sigma_q + \frac{d\ln(a^3)}{dt}  \dot \sigma_q + 
  \left( \frac{q^2}{a^2} + 12 \xi H^2 \right) \sigma_q = 0
\ea
The free vacuum is the Bunch-Davis vacuum defined by
\ba \label{BDmodesigma}   
  \left( i\partial_{\tau} - q  \right)\, 
 \left(a  \sigma_{q}  \right) \to 0
 \qquad {\rm for} \quad \frac{q}{aH} \to \infty 
\ea
In a quasi-de Sitter approximation, the mode functions are
\ba \label{solsigmamin}
  \sigma_{q}(t) &=& \sigma_q^{0\, \rm min}
  \left( 1+ i q\tau \right) e^{-iq\tau}  \, ,
\\
 \label{solsigmaconf}
 \sigma_{q}(t) &=& \sigma_q^{0\, \rm conf}
 \,  e^{-iq\tau}  \, ,
\ea 
for minimally and conformaly coupled scalars where
the normalization constants are 
\ba
\label{normssigma}
  \abs{\sigma_q^{0\, \rm min}}^2 
  =  \frac{H_q^2}{2 q^3}  \, , \qquad 
  \abs{\sigma_q^{0\, \rm conf}}^2 
  =  \frac{1}{2 q}  \, .
\ea

We work at the leading order of the large-${\cal N}$ limit. 
 At this order, 
inspection of the diagrammatic expansion reveals the following 
elements. 
First, gravitational self-interactions are irrelevant. 
It means that the gravitons decouple from the scalar perturbations, and that
scalar self-interactions do not contribute.
Second, 
since each matter loop is enhanced by a factor ${\cal N}$, 
matter fields propagate freely. Therefore 
the only loop corrections to consider 
are matter loops in the two-point function of $\zeta$.
Moreover, only the trilinear vertex $\zeta \sigma \sigma$ contributes to the 
logarithmic part of the one loop correction (the local-regular part
of the loop correction does not contribute to the entropy \cite{paperI},
so that the one-loop correction comming from the vertex 
$\zeta \zeta \sigma \sigma$ need not be considered here).
Finally, an explicit calculation \cite{Weinberg} shows that the relevant part of 
the trilinear vertex responsible for the logarithm is 
\ba \label{Aterm}
  S_{\zeta \sigma \sigma} = - \int\!\!dt \, H_{\rm int}  
  = - \int\!\!dtd^3x \, a^3
 \left( T^{00} + a^2 \delta_{ij}T^{ij}   \right) 
  \left(-\epsilon H a^2 \nabla^{-2} \dot \zeta \right)\, ,
\ea
where $T_{\mu\nu}$ is the energy momentum tensor of matter.  
For minimally and conformaly coupled scalars, the linear combination in \Eqref{Aterm}
is respectively
\ba
\label{clTmin}
  \left(T^{00} + a^2 \delta_{ij}T^{ij}\right)^{\rm min}
  &=& 2 \dot \sigma^2 \, ,\\
\label{clTconf}
  \left(T^{00} + a^2 \delta_{ij}T^{ij}\right)^{\rm conf}  
  &=& \dot \sigma^2  
 + \frac{1}{3}\left( \frac{1}{a^2} (\nabla \sigma)^2 - 2\sigma \ddot \sigma -
   H^2 \sigma^2  \right)       \, . 
\ea

\subsection{Outline of the calculation}

The expectation value of a local (possibly composite) operator $Q(t, {\bf x})$
is given by
 \ba \label{perturb}
   Q(t, {\bf x}) &=&  
   Q_{I}(t, {\bf x})  \, + \,
   i  \,\int_{-\infty}^{t}\!\!dt_2
   \left[ H_{I}(t_2) ,\, Q_{I}(t, {\bf x}) \right]
   \, \nonumber \\
   && - \int_{-\infty}^{t}\!\!dt_2
   \int_{-\infty}^{t_2}\!\!dt_1 \,
   \left[ H_{I}(t_1) , \left[ H_{I}(t_2),\,
   Q_{I}(t, {\bf x}) \right] \right]
   \, + ...
 \ea
The subscript $I$ refers to the interaction picture and will be 
omitted in the following.
In this expression, $H_{I}$ is interacting Hamiltonian, 
e.g. $H_{\rm int}$ in \Eqref{Aterm}. 
The dots stand for higher order corrections and counterterms (whose explicit
form will not be needed here).
The Fourier transform of the one loop correction to covariances 
$\delta {\cal P}_{\xi \xi'}$ can then 
be written
\ba \label{1loop}
   \int\!d^3x \, e^{i{\bf q}{\bf x}}  
 \, \delta\langle \xi(t,{\bf x}) \xi'(t,{\bf 0}) \rangle
  &=& -4 {\cal N} \int\!\!\frac{d^{3}p}{(2\pi)^3} \, \frac{d^3p'}{(2\pi)^3} \, 
 (2\pi)^3\delta^{(3)}\left( {\bf q} + {\bf p} + {\bf p}'  \right)\, 
 \nonumber \\
 && \quad 
 \times \, \int_{-\infty}^{t}\!\!dt_2 V_2  \int_{-\infty}^{t_2}\!\!dt_1 V_1 \, 
 Re\left( {\cal Z}_q^{\xi \xi'} {\cal M}_{pp'}^{\sigma}\right) + ...
\ea
where $\xi$ and $\xi'$ can be either $\zeta$ of $\dot \zeta$, 
and $V(t) = - \epsilon H a^5$.
The same function ${\cal M}$ appears in the calculation of the three covariances. 
It depends only on the matter fields $\sigma_n$ and the loop variables 
$p$, $p'$, $t_1$  and $t_2$.
If the $\sigma$'s are conformally coupled, one finds
\ba \label{Mconf}
  {\cal M}_{pp'}^{\rm conf} &=& 
  \frac{{\cal N}}{36} 
  \frac{1}{a_1^4 a_2^4}\frac{\left(p^2+ p^{'\, 2} - 4pp'\right)^2}{pp'} \, 
  e^{-i(p+p')(\tau_1 - \tau_2)} \, .
\ea
For minimally coupled scalar fields one gets 
\ba \label{Mmin}
  {\cal M}_{pp'}^{\rm min} &=& 
  {\cal N} \, \frac{\left( pp' \right)^2}{a_1^4 a_2^4} \, 
  e^{-i(p+p')(\tau_1 - \tau_2)}  \, .
\ea
Notice that the time dependence of these two expressions is the same.
This remark is essential to understand why the entropy 
from minimally coupled fields is not much larger 
as might have been naively expected 
(since they are not in the conformal vacuum but are parametrically
amplified).
The function ${\cal Z}_q^{\xi \xi'}$ depends only of the
modes $\zeta_q$ and/or $\dot \zeta_q$. 
Using the solution \Eqref{solzeta} of the mode equation, we get 
\sba
 \label{Zqq}
  {\cal Z}_q^{\zeta \zeta} &=& 
  \frac{\abs{\zeta_q^0}^4}{H_1H_2 a_1^2 a_2^2} 
 \left[  e^{-iq(\tau_1 + \tau_2 - 2\tau)}(1-iq\tau)^2 - 
  e^{-iq(\tau_1 - \tau_2)}(1+ q^2\tau^2)\right] \, ,
 \\
  \label{Zqp}
  {\cal Z}_q^{\zeta \dot \zeta} &=& -\frac{q^2}{a^2(t) H}\, 
  \frac{\abs{\zeta_q^0}^4}{H_1H_2 a_1^2 a_2^2} 
 \left[  e^{-iq(\tau_1 + \tau_2 - 2\tau)}(1-iq\tau) - e^{-iq(\tau_1 - \tau_2)}
\right] \, ,
 \\
  \label{Zpp}
  {\cal Z}_q^{\dot \zeta \dot \zeta} &=& \frac{q^4}{a^4(t) H^2}\, 
  \frac{\abs{\zeta_q^0}^4}{H_1H_2 a_1^2 a_2^2} 
 \left[  e^{-iq(\tau_1 + \tau_2 - 2\tau)} - e^{-iq(\tau_1 - \tau_2)}\right] \, .
\sea
In each line, the first term in the brackets 
comes from the term $\langle \zeta_1 \zeta_2 Q \rangle$ of the perturbative 
expansion \Eqref{perturb}, and the second from
$\langle \zeta_1 Q \zeta_2 \rangle$.
It is tempting to dismiss the terms $O(q\tau)$ since for the power spectrum
of super Hubble scales, they represent subleading terms.
These must however be kept to calculate the entropy.
Indeed, we remarked already below Eq. \Eqref{SvN2} that at tree level, 
three powers of $1/q\tau $ cancel 
in the expression of $\det(C)$ so as to ensure that the entropy vanishes, 
see Sec. \ref{sec:Gapprox}.

Then, the key observation is that all the factors of $a$ 
coming from $V(t_{1,2})$, ${\cal Z}$ and ${\cal M}$ cancel 
in the integrand of \Eqref{perturb}. Thus, the dummy variables $\tau_1$ and $\tau_2$
appear only in the phase of exponentials. 
Explicitely, they are the phase factors in the brackets of  
Eqs. (\ref{Zqq}) and from $\cal M$ of Eqs. \Eqref{Mconf} and \Eqref{Mmin}. 
The singular logarithm of the one loop correction 
can be calculated by exchanging the order of integration.
The integration of the first phase in Eqs. (\ref{Zqq}) gives 
\ba \label{timeint1}
 {\cal I}_1(q,p,p') &=& e^{i2q\tau} \, \int_{-\infty}^{\tau}\!\!d\tau_2  \, 
 e^{i(p+p' -q) \tau_2}\int_{-\infty}^{\tau_2}\!\!d\tau_1 \, e^{-i(p+p'+q) \tau_1} 
= - \frac{1}{2q(q+p+p')}\, .
\ea
The remarkable property of this term is its $\tau$-independence.
This is a consequence of the stationarity of de Sitter space, but it 
can be shown that the integral is also finite
in the limit $t\to + \infty$ 
in power law inflation for $\epsilon < 1/3$.
For the second term, we introduce $Q = p + p' + q$, 
and make the change of variables 
$(\tau_2,\tau_1) \mapsto (\tau_2,\delta \tau = \tau_1 - \tau_2)$. We get
\ba \label{timeint2}
    {\cal I}_2(q,p,p') &=& \int_{-T}^{\tau}\!\!d\tau_2  \, \frac{i}{Q+i\epsilon} 
    =\Delta \tau \left[ i {\cal P} \frac{1}{Q} + \pi \delta(Q) \right]\, ,
\ea
where $\Delta \tau = \tau + T$.

The integrals ${\cal I}_1$ and ${\cal I}_2$ times the kernel ${\cal M}$
must then be integrated. Since $p,p',q \geq 0$, 
the Dirac $\delta$ in \Eqref{timeint2} gives a vanishing contribution for
any finite $q$, while the principal value gives a purely imaginary contribution.
Since this second term is only multiplied by real functions in 
Eqs. (\ref{Zqq}), it therefore does not contribute to \Eqref{1loop}.
Hence, only the interfering term 
$\langle \zeta_1 \zeta_2 Q \rangle$ contributes.
Because of the the curvature perturbation couples to matter
via its time derivative $\dot \zeta$, its conjugate momentum 
is not \Eqref{pi_zeta} but $\pi = a^3 \epsilon \dot \zeta/(4\pi G) + 
F(\partial \sigma, \dot \sigma)$ where $F$ is a functional quadratic 
in $\sigma$. These additional terms do not contribute at one loop 
to the logarithmic
part of the variance ${\cal P}_{\zeta \pi}$ and ${\cal P}_{\pi \pi}$ so that
we ignore them. 
Therefore the 1-loop modifications of the variances are
\sba 
  \label{qq}
  \langle \zeta_{\bf q}(t) \zeta_{-{\bf q}}(t) \rangle_{\rm 1-loop} &=& 
  {\cal N} 
  \left( 8 \epsilon_q^2 \abs{\zeta_q^0}^4 \right) \, 
  \left( 1 - \frac{q^2}{a^2H^2} \right) \,{\cal J}\left( \frac{q}{\mu}\right) \, ,
 \\
  \label{qp}
  \frac{1}{2}\langle \left\{ \zeta_{\bf q}(t) ,\,  
  \pi_{-{\bf q}}(t) \right\} \rangle_{\rm 1-loop} &=&  - {\cal N} 
 \left(  8 \epsilon_q^2 \abs{\zeta_q^0}^4 \right) \,
  \left(  - \frac{q^2}{a^2 H}\,\frac{\epsilon a^3 }{4 \pi G}\right) \, 
  {\cal J}\left( \frac{q}{\mu}\right) \, ,
 \\
  \label{pp}
  \langle \pi_{\bf q}(t) ,\,  \pi_{-{\bf q}}(t) \rangle_{\rm 1-loop} &=&  {\cal N} 
 \left(  8 \epsilon_q^2 \abs{\zeta_q^0}^4 \right) \,
  \left(  - \frac{q^2}{a^2 H}  \, \frac{\epsilon a^3}{4 \pi G}  \right)^2 \, 
  {\cal J}\left( \frac{q}{\mu} \right)\, ,
\sea 
where ${\cal J}$ contains the (ultra-violet divergent)
integral over loop momenta. For instance, 
for a minimally coupled field its un-subtracted expression is 
\ba \label{J}
  {\cal J}\left( q \right)  &\equiv& \frac{1}{q}
  \int\!\!\frac{d^{3}p}{(2\pi)^3} \, \frac{d^3p'}{(2\pi)^3} \, 
  (2\pi)^3 \delta^{(3)}\left( {\bf q} + {\bf p} + {\bf p}'  \right)\, 
 \frac{pp'}{q + p + p'}\, .
\ea
The new scale $\mu$ in Eq. \Eqref{qq} 
is an a priori arbitrary scale associated with  
regularization procedure, e.g. dimensional regularization.
The occurrence of $q>0$ in \Eqref{timeint1} 
regulates the momentum integral in the infrared. 
This is the advantage of having inverted the order of the integrals.
Note that the expressions \Eqref{qq} hold for both minimally and conformaly 
coupled fields.
We call $A$ the coefficient of the logarithm of ${\cal J}$.
Its value depends of course on whether $\sigma$ is minimally
or conformally coupled.
It is given in \cite{Weinberg,WeinbergStudent} but we shall not need its 
explicit expression here.  
Taking into account the normalization \Eqref{norms},
the relative change of the power spectrum is 
\ba 
 \frac{ \delta {\cal P}_\zeta }{ {\cal P}_\zeta} = A{\cal N} \epsilon G H^2 
  \ln \left(\frac{q}{\mu}\right)\, ,
\ea
up to terms $(q/aH)^2$.

The covariance matrix 
does not grow with a certain power of $a$ as it could have done a priori. 
It is given by 
\ba
  \det(C) = \frac{1}{4} + O\left( \epsilon_q  G H_q^2{\cal N}  A 
 \ln(q/\mu) \right)^2\, .
\ea
In conclusion at one loop approximation, 
\ba
  S = O(\epsilon_q \times 10^{-10} \times {\cal N})^2 \, ,
\ea
even though the power spectrum is modified to linear order in 
$\epsilon_q \times G H^2 \times {\cal N}$.
Comparing with the entropy of a separable state in the 
 radiation dominated era $S_{\rm sep} = 100 \ln(10)$,
the reduced density matrices remain entangled (not separable) 
at the end of inflation, 
unless 
the number of fields is larger than $1/\epsilon_q \times 10^{10} $
but in this case the whole perturbative treatment is no longer valid.

The reduced density matrix of modes of opposite wave vectors
remains very pure, and can 
thus be interpreted as a "dressed squeezed vacuum state".
This is an example of the case discussed in Sec. \ref{sec:heuristic}.
We stress that this conclusion could not be expected {\it a priori} on 
the basis of an analogy with field theories in the Minkowski vacuum.
In inflation, matter fluctuations of non conformal fields 
are parametrically amplified just as the curvature perturbations,
so that they could have been responsible for a strong decoherence. 
As we saw, this is not the case and  
 mainly follows from the fact that 
matter couples to $\zeta$ via its energy-momentum tensor, 
see \Eqref{Aterm}.

\section{Conclusion and outlook}

We found that in multi-field inflation scenarios, 
the entropy grows at a high rate,
typically $dS/dN = 6$, as long as some isocurvature mode
decay polyniomaly in the number of efolds after horizon exit.
By contrast, in single field inflation we found (at one-loop) no evidence
for decoherence, so that the state remains essentially pure. 
Intermediate rates between these two cases seem hard to find 
because of the efficiency of the mechanism of parametric amplification
(in amplifying or canceling the action of a noise).

 We concentrated on the entropy of curvature perturbations, 
 but our method may as well be applied to tensor modes. 
 We conclude on the perspectives to apply this method to the regimes of 
 (p)reheating and horizon reentry. 
 One should distinguish between sub- and superhorizon scales.
 The evolution of the former is described by 
 a regime of broad parametric resonance. This is 
 a process characterized by its efficiency. Decoherence should therefore 
 be achieved very rapidely.
 The evolution of super-Hubble modes depends on whether isocurvature perturbations
 are excited or not. 
 The discussion of this case can be incorporated in the model
 of two field inflation of Sec \ref{sec:iso} 
(see \cite{reheating} as well as references therein).

 After reheating but before decoupling, 
 it is expected that the coupling of curvature modes 
 entering the horizon with the radiation 
 density perturbations in quasi-thermal equilibrium  
 will erase the possible remaining 
 quantum features after typically one oscillation
 (this should not be mistaken with the thermalization of the 
 perturbations which occurs much later).
 We have little to say about this regime of horizon reentry. A field theoretic
 proof remains a formidable task because long wavelength radiation perturbations 
 are not thermal fluctuations since we observe acoustic oscillations. 
 Since this phenomenon is reliably described by a hydrodynamic model, it
 means that scales decouple. As a result, we expect that the 
 curvature perturbations entering the horizon dominantly couple 
 to the same scales of the plasma.
 A proper investigation of the regime of horizon reentry therefore requires first
 to write an effective 
 field theoretic model of these out-of-equilibrium long wavelength modes.
 This is arguably an academic question, since we saw in \cite{paperI} that
 the loss of quantum coherence 
 occurs on time scales much shorter than the characteristic 
 time of thermalization.

\acknowledgments{The work of D.C. is supported by the Alfried Krupp Prize for Young 
University Teachers of the Alfried Krupp von Bohlen und Halbach
Foundation.}

\begin{appendix}

\section{Derivation of Eq. \Eqref{backreactEOM}}

\subsection{Equation for the anti-commutator}
\label{app:CTP}

When the 
Hamiltonian depends explicitly on time, there is 
no stable ground state. In this case, Green functions, e.g.
\ba \label{exactGreen}
   G(x,y) = \bra{0 {\rm in}} {\cal T} \varphi(x)
   \varphi(y) \ket{0 {\rm in}} \, ,
\ea 
are expectation values in the 'in'-vacuum $\ket{0 {\rm in}}$
which cannot be expressed in terms 
of Feynman graphs with internal lines 
corresponding to time-ordered propagators.
Rather, their generating functional is 
given by the transition amplitude of two 
'in'-vacua in the presence of external sources $J_+$ and $J_-$
(see for instance \cite{JordanHuCalzeta}),
 \ba \label{Winin}
   e^{iW[J^+,\,J^-;\rho_{\rm in}]} 
   &=& {\rm Tr} \left\{ 
   {\cal T} e^{i\int_{-\infty}^{t_{out}}dt\, J_+(x) \varphi
   (x)} 
   \,  \rho_{\rm in} \, 
   \widetilde {\cal T} e^{-i\int_{-\infty}^{t_{out}}dt\, 
   J_-(x) \varphi
   (x)}  
   \right\}\, ,
\ea
where $\widetilde {\cal T}$ is the reversed-time ordered product and 
$J_+$ and $J_-$ are the two classical sources associated with the 
two branches of evolution, forward and backward in time respectively.  
Note that the operation of taking the trace
couples the forward and backward time evolutions. 
As a result, $W[J_+,J_-]$ generates four types of connected 
two-point functions: the time-ordered propagator
\ba \label{G++}
   G_{++}(x,y) &\equiv& i\langle {\cal T} \varphi(x) \varphi(y)  \rangle = 
  \frac{\delta W}{\delta J_+(x) \delta J_+(y)}\vert_{J_+ = J_- =0}    \, ,
\ea
the reverse times ordered propagator
\ba\label{G--}
  G_{--}(x,y) &\equiv& i\langle \widetilde{\cal T} \varphi(x) \varphi(y)  \rangle =
  \frac{\delta W}{\delta J_-(x) \delta J_-(y)}\vert_{J_+ = J_- =0}  \, ,
\ea 
and the two on-shell two-point functions
\ba\label{G-+}
   G_{-+}(x,y) &\equiv& i\langle \varphi(x) \varphi(y)  \rangle = 
  \frac{\delta W}{\delta J_-(x) \delta J_+(y)}\vert_{J_+ = J_- =0}  \,,
\\
 \label{G+-}
   G_{+-}(x,y) &\equiv& i\langle \varphi(y) \varphi(x)  \rangle = 
  \frac{\delta W}{\delta J_+(x) \delta J_-(y)}\vert_{J_+ = J_- =0}    \, .
\ea
The latters are not time ordered because
they are build from operators coming from different branches of
 time evolution. 
Eqs. (\ref{G++}-\ref{G+-}) are easily derived using the second equality in
\Eqref{Winin}.
 
The path integral representation of \Eqref{Winin} is
\ba
   e^{iW[J^+,\,J^-;\rho_{\rm in}]}    
   &=& 
   \int\!\!d\varphi_\Sigma^- \, d\varphi_\Sigma^+ \, 
   \delta\left( \varphi_\Sigma^- - \varphi_\Sigma^+\right) \,
   \int\!\!d\varphi_\infty^- \, d\varphi_\infty^+ \,   
   \bra{\varphi_{\infty}^+} \rho_{\rm in} \ket{\varphi_{\infty}^-} \,
\nonumber\\
  &&\times \,
 \int\!\!{\cal D}\phi^+ {\cal D}\phi^-
  e^{\left( i{\cal S}[\varphi_\infty^{+},\,\varphi_\Sigma^{+}](\phi^+)
   + iJ^+ \phi^+
   - i{\cal S}[\varphi_\infty^{-},\,\varphi_\Sigma^{-}](\phi^-)
   - iJ^- \phi^- \right)}\, ,
\qquad 
\ea
 where
${\cal S}[\varphi_\infty^{+},\,\varphi_\Sigma^{+}](\phi^+)$ 
 is the classical action evaluated for the paths
 $\phi^+(z^0,{\bf z})$ with fixed end-points
 $\phi^+(\infty,{\bf x}) = \varphi_{\bf x}^+$ and
 $\phi^+(\Sigma,{\bf y}) = \varphi_{\bf y}^+$.

In the above equations, $\varphi$ denotes the collection of fields. 
In cosmological settings, we decompose these fields into
the adiabatic perturbations $\zeta$ and the rest we call $\sigma$.
Then, the so-called influence-functional (IF) encodes the effective 
dynamics of $\zeta$ when these extra 
fields have been integrated: 
 \ba
    e^{i S_{IF}[\zeta_+, \, \zeta_-]} &\equiv&
    \int_{-\infty}^{+\infty}\!\!d\sigma^+_\Sigma d\sigma_\Sigma^- \, 
   \delta[\sigma_\Sigma^+ -\sigma_\Sigma^-]
    \int_{-\infty}^{+\infty}\!\!d\sigma_\infty^+ d\sigma_\infty^- \, \,
     \bra{\sigma_{\infty}^+} \rho^\sigma_{\rm in} \ket{\sigma_{\infty}^-} \,
    \nonumber \\
   && \qquad \times \, \int\!\!{\cal D}\sigma^+ {\cal D}\sigma^-
   \, e^{i\left( S[\sigma_+] - S[\sigma_-] +
   S_{int}[\sigma_+,\, \varphi_+] - S_{int}[\sigma_-,\, \zeta_-] \right)}\, .
 \ea
 We can evaluate perturbatively the IF by expanding the terms
 $e^{i(S_{int}[\Phi_+] - S_{int}[\Phi_-])}$ to a 
 given order in $\hbar$, then by
 taking the expectation value
 for the field $\sigma$, and finally
 by re-exponentiating the result.
 Hence we get
 \ba \label{quadapproxS_{IF}}
   S_{IF}[\zeta_-, \, \zeta_+] &=&
   i \left(\langle  S_{int}[\zeta_+]\rangle -
   \langle S_{int}[\zeta_-] \rangle\right)
\nonumber \\
  &-&
  \frac{1}{2} \left\{
   \, \langle S_{int}[\zeta_+] S_{int}[\zeta_+] \rangle_{\rm con}
   + \langle S_{int}[\zeta_-]  S_{int}[ 
   \zeta_-] \rangle_{\rm con}
   \right.
   \nonumber \\     
 &&  \left.\quad
 - \langle S_{int}[\zeta_+] S_{int}[\zeta_-] \rangle_{\rm con}
   - \langle S_{int}[\zeta_-] S_{int}[\zeta_+] \rangle_{\rm con}
  \, \right\} 
+ ... \qquad
 \ea
 where $\langle \, \,\, \rangle_{\rm con}$ means the
 (connected part of) expectation value in $\rho^\sigma_{\rm in}$, 
 the initial state of
 the field $\sigma$.

The Gaussian approximation consists in keeping 
only the part quadratic in
$\zeta_\pm$ of this functional. Then the
 generating functional is given by
 \ba \label{Zctp}
   e^{iW_{\rm gauss}[J^+,\,J^-;\rho_{\rm in}]} 
   &=& \int\!\!{\cal D}\zeta^+ {\cal D}\zeta^-
   \exp\{-\frac{1}{2} \,\, {}^t\!Z\, {\cal M} Z + {}^t\!{\cal J} Z\}
   \nonumber\\
&\propto& 
\exp\left( - \frac{1}{2} {}^t\!{\cal J} {\cal M}^{-1} {\cal J} \right)\, ,
 \ea
 where ${}^t Z = \left(\zeta_+ \,\, \zeta_- \right)$
 and ${}^t {\cal J} = i\left(J_+ \, , -J_-\right)$.
The quadratic form $\cal M$ is by definition 
\ba
  \frac{1}{2} \,\, {}^t\!Z \, {\cal M} Z = 
  {\cal S}[\zeta_+] - {\cal S}[\zeta_-]
  + S_{\rm IF}^{\rm gauss}[\zeta_+, \zeta_-]\, .
\ea
The action $\cal S$ is given by eq. \Eqref{Squadzeta}
whereas the Gaussian approximation of the IF is
\ba \label{canonicalSeff}
  S_{\rm IF}^{\rm gauss} = \frac{1}{8\pi G} \int\!\!dt d^3x (a^3(t)\epsilon) 
  \int\!\!dt'd^3y (a^3(t')\epsilon)\,  
  \zeta_a(x) \Sigma_{ab}(x,y) \zeta_b(y)\, ,
\ea
where the indices $a,b=\pm$.
This equation 
defines the self-energy matrix $\Sigma_{ab}$.
We have introduced twice the factor $a^3\epsilon$,
which it is already present in the kinetic action (\ref{Squadzeta}),
because it simplifies the forthcoming equations.

Taking the functional derivatives with respect to the sources, 
we arrive at a 
system of two linear coupled equations. 
The system decouples when using  
the (odd) commutator (also called the spectral function)
 $\rho = i(G_{-+} - G_{+-})$
and the (even) anticommutator 
$G = (G_{++} + G_{--})/2 = (G_{+-} + G_{-+})/2$.
Indeed, 
using the definition of the free propagators
$\delta_2 {\cal S} = {\cal D}_x = G_0^{-1}$, one gets
\ba \label{eqnrho}
   &&{\cal D}_x\rho(x,y) + 
  \int_{y^0}^{x^0}\!\!d^4z\,  D(x,z)\rho(z,y) = 0 \, ,
\\ 
 \label{eqnG}
&&{\cal D}_x G(x,y) + 
  \int_{-\infty}^{x^0}\!\!d^4z\,  D(x,z)G(z,y) = 
  \int_{-\infty}^{y^0}\!\!d^4z\,  N(x,z) \rho(z,y)   \, .
\ea
Straightforward algebra gives the odd and the even kernels 
\ba \label{NetD}
   D(x,x') &=& i a^3(t') 
  \left[ \Sigma_{-+}(x,x') - \Sigma_{+-}(x,x')  \right]\, ,
\nonumber \\
N(x,x') &=& \frac{a^3(t')\epsilon}{2} 
  \left[ \Sigma_{-+}(x,x') + \Sigma_{+-}(x,x')  \right]\, ,
\ea
in terms of $\Sigma_{ab}$, the self-energy matrix of $\zeta$.

We conclude this part by a comment.
In the Gaussian approximation it is 
always possible to write the effective action 
with only the field (here $\zeta$)
but no field derivatives. 
Should one find, after computation of the effective action, field derivatives
(e.g. because in the total action 
the curvature perturbation couples to other fields 
via $\dot \zeta$), one would simply do an integration by parts in order 
to bring the effective action into the form \Eqref{canonicalSeff}.
This operation is a linear canonical transformation which  
changes neither the equations (\ref{eqnrho},\, \ref{eqnG})
nor the value of the determinant of the covariance matrix \cite{paperI}.
Moreover, with the action written in the form \Eqref{canonicalSeff},
the canonical momentum is still given by
$\pi = ({a^3\epsilon}/{4\pi G})\, \dot \zeta$
which justifies the identities \Eqref{defrhored}.

\subsection{The effective quantum source}
\label{app:trick}

Since the unknown of Eqs. (\ref{eqnrho}) and (\ref{eqnG}) are real functions, they are apt for a numerical analysis. 
Alternately, for an analytical treatment it is simpler 
to go one step backwards and consider the 
operator $\zeta_{\bf q}$ coupled to a
quantum mechanical source $\xi_{\bf q}$ 
whose statistical properties
are such that the anti-commutator of $\zeta_{\bf q}$ 
obeys by construction Eq. (\ref{eqnG}).
This effective source is 
not just an artificial trick, because it coincides with
the true fluctuating source operator when 
non-Gaussianities are neglected,
as it is the case for the two models considered in the
body of the paper.

Let us consider the following Heisenberg equation
\ba \label{ximode} 
  {\cal D}_t \Phi_{\bf q}(t) + 
  \int_{-\infty}^{t}\!\!ds\, 
  D(t,s,{q}) \, \Phi_{\bf q}(s) = 
  \xi_{\bf q}(t) \, ,
\ea
where $\Phi_{\bf q}$ and  $\xi_{\bf q}$ are two operators.
We also introduce the anti-commutator 
\ba
   {\cal G}_{q}(t,t') \, \delta^3({\bf q}-{\bf q'}) 
   = \frac{1}{2} {\rm Tr} \left[ \rho_\Phi \, \rho_\xi \,  
 \left\{  \Phi_{\bf q}(t), \,  \Phi_{-\bf q'}^\dagger(t') 
   \right\}
  \right]\, . 
\ea
For an appropriately chosen anti-commutator of $\xi_{\bf q}$,
we now show that given some initial conditions, 
the function $ {\cal G}_{q}(t,t')$ 
solves for the (Fourier transform
in space) of \Eqref{eqnG} with the same initial conditions.
Hence it is equal to the anti-commutator $G(t,t',q)$.

To determine the statistical properties of the noise $\xi_{\bf q}(t)$ that 
reproduce the action of the r.h.s. of \Eqref{eqnG}, 
we let the integro-differential operator on the l.h.s. of \Eqref{eqnG} 
act on ${\cal G}_q$, and calculate the result using \Eqref{ximode}. 
Since the action of taking
the trace commutes with the partial derivation and integration, we have
\ba \label{intermed2}
 \delta^3({\bf q}-{\bf q'}) \, 
 \left[ {\cal D}_t  {\cal G}_{q}(t,t') + 
  \int_{-\infty}^{t}\!\!ds\,  D(t,s,{q}) {\cal G}_{q}(t,t') \right] = 
  \langle  \left\{
  \xi_{\bf q}(t), \,  \Phi_{-\bf q'}^\dagger(t') \right\}
  \rangle \, .
\ea
To calculate the anticummutator on the r.h.s., we need the solution of 
\Eqref{ximode}. Remembering that the exact retarded Green function of 
$\Phi_{\bf q}$ is the spectral function multiplied by a theta function, i.e.  
$\theta(t-t') \rho_q(t,t')$, the general solution of \Eqref{ximode} is
\ba
  \Phi_{\bf q}(t) =  \Phi_{\bf q}^0(t) + \int_{-\infty}^{t}\!\!ds \, 
  \rho_{q}(t,s) \,  \xi_{\bf q}(s) 
\ea
where $\Phi_{\bf q}^0(t)$ is the solution of the homogeneous equation.
Since it is independent of $\xi(t)$, upon 
substitution of this solution into the r.h.s. of \Eqref{intermed2} one finds,
\ba
  && \delta^3({\bf q}-{\bf q'}) \,  
  \left[{\cal D}_t {\cal G}_{\bf q}(t,t') + 
  \int_{-\infty}^{t}\!\!ds\, 
  D(t,s,{q}) {\cal G}_{\bf q}(t,t')\right] = 
 \nonumber \\  
  && \qquad \qquad \qquad 
  \frac{1}{2} \int_{-\infty}^{t'}\!\!ds  \, 
  \langle  \left\{
  \xi_{\bf q}(t), \,  \xi_{-\bf q'}(s)   \right\}
  \rangle 
\, \rho_{q}(s,t')
\ea
Identification with \Eqref{eqnG} finally yields
\ba \label{noisexi}
  \frac{1}{2} \langle  \left\{ \xi_{\bf q}(t) , \, \xi_{-\bf q'}(s) \right\}
  \rangle = N(t,s, q)\delta^3({\bf q}-{\bf q'}) \, \, 
\ea
where $N(t,s,q)$ is the (Fourier transform of the) 
kernel appearing in Eq. \Eqref{eqnG}.
In conclusion, the dynamics of $\zeta$ 
is equivalent to the
"open dynamics" \Eqref{ximode} of 
$\Phi$ with the same spectral function $\rho$ and 
subject to the Gaussian 
source $\xi$ characterized by the spectrum 
\Eqref{noisexi}. 

This result is general because we are only considering the two-point functions. Because of the linearity of the equations
and the Gaussianity of the noise, this approach 
can be also phrased in terms of a stochastic
(commutating) c-number source $\xi_{\bf q}$,
 see the stochastic approach
of quantum gravity in \cite{HuVerda}.
This equivalence is used in Sec. \ref{sec:expdecayingcorrections}.

Finally, 
from the definitions \Eqref{defrhored} and the identification $G = {\cal G}$, 
the covariances can always be written as a the sum
of a free (and decaying) part and a driven part:
\sba \label{covstoch}
   {\cal P}_{\zeta \zeta} &=& \langle \Phi_{\bf q}^0(t)\Phi_{\bf q}^0(t) \rangle 
  + \int\!\!dt_1 dt_2 \,
   G_{\rm ret}(t,t_1)G_{\rm ret}(t,t_2) N(t_1,t_2) \, ,
\\
   {\cal P}_{\zeta \dot \zeta} &=& 
  \frac{1}{2} \langle \left\{ \Phi_{\bf q}^0(t) , \, \dot \Phi_{\bf q}^0(t) \right\} \rangle 
  + \int\!\!dt_1 dt_2 \, G_{\rm ret}(t,t_1) \partial_t G_{\rm ret}(t,t_2) N(t_1,t_2)    \, ,
\\
   {\cal P}_{\dot \zeta \dot \zeta} &=& 
  \langle  \dot \Phi_{\bf q}^0(t)  \dot \Phi_{\bf q}^0(t) \rangle 
  + \int\!\!dt_1 dt_2 \, \partial_t G_{\rm ret}(t,t_1) 
  \partial_t G_{\rm ret}(t,t_2) N(t_1,t_2)\, .
\sea

\section{Intermediate results for Sec. \ref{sec:iso}}
\label{app:longlambda}

\subsection{The free retarded propagator in the long wavelength limit} 
\label{app:Gret}

The free retarded propagator of \Eqref{EOMzetafree} is
\ba
  G_{\rm ret}(t,t',q) &=&  2\theta(t-t') 
  \frac{\zeta_d(t) \zeta_g(t') - \zeta_g(t) \zeta_d(t')}{W(t')}
\ea
where $\zeta_g$ and $\zeta_d$ are the homogeneous growing and decaying solutions of 
\Eqref{EOMzeta+iso}, and $W(t)$ is their Wronskian.
For our calculation, we only need to retain the leading terms 
of their expansion in powers of $(q/aH)^2$.
In the limit $q\to 0$, the Wronskian solves for the equation
\ba
  \dot W + \frac{d \ln(a^3\epsilon)}{dt} W = 0
\ea
whose solution can be conveniently written as 
\ba \label{wronskian}
  W(t) = W(t_*) \frac{a_*^3 \epsilon_*}{a^3(t) \epsilon(t)}
\ea
where $t_*$ is the time of horizon crossing $q=a_* H_*$.
The solution of \Eqref{EOMzetafree} is
\ba \label{modesol}
  \zeta_q(t) = \zeta_q^0 \left[ 1 + O\left( \frac{q^2}{a^2 H^2} \right) \right] 
  + A_q\left[ \int_{t}^{\infty}\!\!\frac{dt'}{a^3(t') \epsilon(t')} + 
  O\left( \frac{q^2}{a^2 H^2} \right)\right]
\ea
The coefficients $\zeta_q^0$ and $A_q$ of the growing and decaying solutions are
related by the Wronskian condition
\ba 
  i\left( \zeta_q \dot \zeta_q^* - c.c \right) = 
  \frac{2{\rm Im}\left( \zeta_q^{0 \, *} A_q\right) }{a^3(t) \epsilon(t)} = 
  \frac{4\pi G}{a^3(t) \epsilon(t)}
\ea
Combining these results, we get
\ba \label{Gret}
  G_{\rm ret}(t,t',q) &=& 
  \theta(t-t')a^3(t') \epsilon(t')\int_{t}^{t'}\!\!
  \frac{dt_1}{a^3(t_1) \epsilon(t_1)} +  
  O\left( \frac{q^2}{a^2 H^2} \right)
 \nonumber \\
  &\simeq& \frac{\theta(t - t' )}{3H} 
  \left[ 1-  \left(\frac{a(t')}{a(t)}  \right)^3   \right] \simeq  
  \frac{\theta(t - t' )}{3H}   
\ea
To get the second line, we substituted the solution \Eqref{modesol} 
of the mode equation 
in the slow-roll approximation.
We also give the first time derivative of $G_{\rm ret}$ in that approximation,
\ba \label{partialGret}
  \partial_t G_{\rm ret}(t,t',q) 
  &\simeq& \theta(t-t') \left( \frac{a(t')}{a(t)}  \right)^3
\ea

\subsection{The covariances of Sec. \ref{sec:strongcoupling}}
\label{app:xi-model}

We use the solution \Eqref{solzeta} of the mode equation. 
The retarded Green function is
\ba
   G_{\rm ret}(t,t_1) &=& -2\theta(t-t_1) \, 
  \frac{Im\left\{ \zeta_q(t) \zeta_q^*(t_1) \right\}}{W(t_1)}
\nonumber \\
   &=& \frac{\theta(t-t_1)}{H^2}  \, Im\left\{ (1-ix) (1+ix_1) e^{i(x-x_1)}  \right\}
\ea
We use again the variable 
\ba
   x = \frac{q}{aH} = - q\tau
\ea
The noise kernel is
\ba
   N(t_1,t_2) = 9H^2 g^2 {\cal P}_0   \qquad 
   {\rm for}\quad  t_\xi \geq t_1,t_2 \geq t_*
\ea

The covariances calculated from \Eqref{covstoch} are given by
\sba \label{exactcoviso}
   {\cal P}_{\zeta \zeta} &=& {\cal P}_0 \left\{ 1 + x^2 + g^2 f^2(x;x_\xi)    \right\}
\\
   {\cal P}_{\zeta \dot \zeta} &=& 
  -Hx^2{\cal P}_0 \left\{ 1 -  g^2 f(x;x_\xi)h(x;x_\xi)    \right\}
\\
   {\cal P}_{\dot \zeta \dot \zeta} &=&  H^2x^4
  {\cal P}_0 \left\{ 1 + g^2 h^2(x;x_\xi)    \right\} 
\sea
and the determinant of the covariance matrix is
\ba \label{detCtoymodel}
   \det(C) = \frac{1}{4x^2} \left\{ x^2 \left[ 1 + g^2h^2(x;x_\xi)   \right]  
  + g^2 \left( f-h   \right)^2  \right\}
\ea
The functions $f$ and $h$ come from the integral of $G_{\rm ret}(t,t_1)$
and $\partial_t G_{\rm ret}(t,t_1)$ respectively. Their expressions are
\ba
   f(x;x_\xi) &=& Im\left\{ (1-ix)e^{ix} {\cal J}(x;x_\xi)   \right\} = 
   h(x;x_\xi) - x \,Re\left\{ e^{ix} {\cal J}(x;x_\xi)   \right\}
\\
   h(x;x_\xi) &=& Im\left\{ e^{ix} {\cal J}(x;x_\xi)   \right\}
\ea 
and the function ${\cal J}(x;x_\xi)$ is
\ba
   {\cal J}(x;x_\xi) &=& 3\int_{1}^{x}\!\!\frac{dx_1}{x_1^4} \, \theta(x_\xi -x) \,
   (1+ix_1)e^{-ix_1}
\ea

\subsubsection{Entropy growth for $t \leq t_\xi$}

In that case
\ba
   {\cal J}(x;x_\xi) &=& -i{\cal L}(x) - e^{-ix}\left( \frac{1}{x^3} 
 + \frac{i}{x^2} + \frac{1}{x} \right) + e^{-i} \left(  
  2+i\right)
\ea
where the first term is a logarithm
\ba
  {\cal L}(x) = \int_{1}^{x}\!\!\frac{dx_1}{x_1} \, e^{-ix_1}
  = E_1(i) - E_1(ix) = \ln(x)  + O(x)
\ea
Combining these expressions, we get
\ba
    f  &=&  - \ln(x) + C_1 + O(x) 
\nonumber \\
    h &=& -\frac{1}{x^2} - \ln(x) + C_2 +O(x)
\ea
from which we obtain the expressions (\ref{C11iso}-\ref{C22iso}) 
as well as 
\ba
   \det(C) = \frac{g^2}{4x^6} \left\{ 1 + O(x^2)  \right\} 
\ea

\subsubsection{Constant entropy after $t_\xi$}
\label{app:verif}

We see on \Eqref{detCtoymodel} that the entropy is constant provided
\ba \label{condition}
  x^2 h^2 + \left( f-h   \right)^2 = \lambda x^2
\ea
where $\lambda$ is a constant.
For $t \geq t_\xi$, the integral ${\cal J}$ is equal to its value at $t_\xi$
\ba 
  {\cal J}(x;x_\xi) = {\rm cte} = {\cal J}_\xi \equiv 
  \vert {\cal J}_\xi \vert e^{i\varphi}
\ea
Substituting into the expressions of $f$ and $h$, we find
\ba
   f - h &=& - x Re\left\{ e^{ix} {\cal J}_\xi \right\} = 
  - x \vert {\cal J}_\xi\vert  \cos(x+\varphi)
\nonumber  \\
   h &=& x Im\left\{ e^{ix} {\cal J}_\xi \right\} = 
  x \vert {\cal J}_\xi \vert \sin(x+\varphi)
\ea
The condition \Eqref{condition} is realized and we have
\ba
   \det(C) = \frac{1}{4} \left\{ 1 + g^2 \vert {\cal J}_\xi \vert^2  \right\}
\ea
with $\vert {\cal J}_\xi \vert = x_\xi^{-3}$.

One also checks that the covariances \Eqref{exactcoviso} verify
the identities \Eqref{identity1} and \Eqref{identity2} toghether 
with the {\it free} equation of motion.

\end{appendix}

\end{document}